\documentclass{aa} 
\usepackage{graphicx}
\usepackage[normalem]{ulem}
\usepackage{subcaption}
\usepackage{amssymb}
\usepackage{natbib}
 \usepackage{tabularx}
\usepackage{appendix}
\usepackage{enumerate}
\usepackage{threeparttable}
\usepackage[colorlinks, citecolor=blue, linkcolor=blue, urlcolor=blue]{hyperref}
\usepackage{etoolbox}
\usepackage[figuresright]{rotating}

\begin{document}
\title{An X-ray selected catalog of extended galaxy clusters from the {\em{ROSAT}} All-Sky Survey (RXGCC)}
\author{Weiwei Xu\inst{1,2}
\and
Miriam E. Ramos-Ceja\inst{3}
\and
Florian Pacaud\inst{2}
\and
Thomas H. Reiprich\inst{2}
\and
Thomas Erben\inst{2}
}
\institute{
The Kavli Institute for Astronomy and Astrophysics at Peking University (KIAA-PKU), Peking University, Yiheyuan Road 5, 100871 Beijing, China (\email{wwxu@pku.edu.cn})
\and
Argelander-Institut f\"ur Astronomie (AIfA), Universit\"at Bonn, Auf dem H\"ugel 71, 53121 Bonn, Germany
\and
Max-Planck-Institut f{\"u}r extraterrestrische Physik, Giessenbachstra{\ss{}}e, 85748 Garching, Germany
}
\date{Received xx, 202x; accepted xx, 202x}

\abstract
{There is a known tension between cosmological parameter constraints obtained 
from the primary cosmic microwave background (CMB) and those from 
galaxy cluster samples. One possible explanation for this discrepancy could be that the incompleteness of 
detected clusters is higher than estimated, and certain types of groups or 
clusters of galaxy have been missed in the past.
}
{We aim to search for galaxy groups and clusters with particularly extended surface brightness distributions, by creating a new X-ray selected catalog of 
extended galaxy clusters from the ROSAT All-Sky Survey (RASS),
using a dedicated source detection and characterization algorithm optimized for extended sources.}
{Our state-of-the-art algorithm includes multi-resolution filtering, 
source detection and characterization. Through extensive 
simulations, the detection efficiency and sample purity are investigated. 
Previous cluster catalogs in X-ray and other wave-bands, as well as 
spectroscopic and photometric redshifts of galaxies are used for the cluster
identification.}
{We report
a catalog of galaxy clusters at high galactic latitude based on the ROSAT 
All-sky Survey, named as RASS-based extended X-ray Galaxy Cluster 
Catalog (RXGCC), which includes $944$ groups and clusters. Out of this number, 
$641$ clusters have been identified through intra-cluster medium (ICM) emission previously ($Bronze$), $154$ known optical and infrared clusters are detected as X-ray clusters for the first time ($Silver$), and $149$ identified as clusters for the first time ($Gold$). Based on $200$ simulations, the contamination ratio of the detections which were identified as clusters by ICM emission, and the detections which were identified as optical and infrared clusters in previous work is $0.008$ and $0.100$, respectively.   
Compared with $Bronze$ sample, the $Gold+Silver$ sample is less luminous, less massive, and has a flatter surface brightness profile. Specifically, the median flux in [$0.1-2.4$]~keV band for $Gold+Silver$ and $Bronze$ sample is $2.496\times 10^{-12}~$erg/s/cm$^2$ and $4.955\times 10^{-12}$~erg/s/cm$^2$, respectively. The median value of $\beta$ (the slope of cluster surface brightness profile) is $0.76$ and $0.83$ for $Gold+Silver$ and $Bronze$ sample, respectively. This whole sample is available at \url{https://github.com/wwxu/rxgcc.github.io/blob/master/table_rxgcc.fits}.
}

\titlerunning{An X-ray detected sample of galaxy clusters from RASS}
\authorrunning{W. Xu et al.}

\maketitle

\section{Introduction}
\label{sec:introduction}

Among the different cosmological probes, galaxy clusters are 
one of the most important ones to test the standard cosmological 
scenarios \citep{Dunkley2009, Kowalski2008, Reiprich2002, Seljak2002, Dahle2006, Pedersen2007, Rines2007, Wen2010, Allen2008, Allen2011, Abbott2018_gc}. 
As the largest gravitational systems in the universe, their 
spatial distribution and number density are sensitive to the dark matter and dark
energy content \citep[e.g.,][]{Borgani2001, Reiprich2002, Seljak2002, Viana2002,Kravtsov2012,Schellenberger2017}. However, cosmological constraints derived from the primary cosmic microwave
background \citep[CMB,][]{
Planck2016a,Planck2018_overview,Planck2018_cosmo_para,Abbott2020} and those from galaxy cluster studies are in tension \citep[see, e.g., Fig.~17 in][]{2019SSRv..215...25P}. 
One of the possible explanations is the under-estimation of the incompleteness of detected clusters.

The intra-cluster medium (ICM) allows us to detect galaxy clusters using two different wavelengths: X-ray and sub-millimetre bands. Compared with cluster observation in other bands, such as in optical wavelength, the ICM observation trace genuine deep gravitational potential wells, and is less affected by projection effects. The X-ray emission includes the bremsstrahlung and line emission \citep[e.g.,][]{2013SSRv..177..195R}, while sub-millimeter emission comes from the Sunyaev–Zel$'$dovich effect \citep[][]{Sunyaev1970,Sunyaev1972,Sunyaev1980}.

Kinds of methods are used to detect galaxy clusters in X-ray data. The most widely used are the sliding-cell algorithm \citep[e.g.][]{Voges1999}, the Voronoi tesselation and percolation method (VTP, \citealt{Ebeling1993, Scharf1997, Perlman2002}), and the wavelet transformation \citep{Rosati1995, Vikhlinin1998, Pacaud2006, Mullis2003, Lloyd-Davies2011, Xu2018}. These methods have distinct advantages and disadvantages of cluster detections \citep[e.g.][]{Valtchanov2001}.

Up to the mid-2019, ROSAT was the only X-ray telescope performed an all-sky imaging survey \citep[RASS,][]{Truemper1992,Truemper1993} in the $0.1-2.4$~keV energy band. 
There are several ROSAT-based cluster catalogs, 
some of the most important are: REFLEX (ROSAT-ESO Flux Limited X-ray Galaxy Cluster Survey, \citealt{Bohringer2004}), NORAS (Northern ROSAT All-Sky galaxy cluster survey, \citealt{Bohringer2000, Bohringer2017}), BCS (ROSAT Brightest Cluster Sample, \citealt{Ebeling1998, Ebeling2000}), SGP (a Catalog of Clusters of Galaxies in a Region of 1 steradian around the South Galactic Pole, \citealt{Cruddace2002}), NEP (the ROSAT North Ecliptic Pole survey, \citealt{Henry2006}), and MACS (Massive Cluster Survey, \citealt{Ebeling2001}). These catalogues, and few more, have been compiled in the so-called MCXC catalogue (Meta-Catalogue of X-ray detected Clusters, \citealt{Piffaretti2011}). 
Besides ROSAT, XMM-Newton and Chandra observatories are also widely used to identify galaxy clusters. With better resolution and sensitivity, and longer exposure times, they could identify faint clusters at high redshift, such as the works of \citet{Mehrtens2012}, \citet{Clerc2012}, \citet{Clerc2016}, \citet{Takey2011,Takey2013,Takey2014,Takey2016}, \citet{Pacaud2016}, to list a few. 

As the successor of ROSAT, the eROSITA X-ray satellite (extended Roentgen Survey with an Imaging Telescope Array, \citealt{Merloni2012}) is scanning the whole sky in the $0.1-10$~keV energy band. eROSITA will complete $8$ scannings in its first $4$~years, and reach a sensitivity of $25$ times better\footnote{It refers to the flux limit for point sources, which is $5\times 10^{-13}$ erg/s/cm$^2$ and $1.1\times 10^{-14}$ erg/s/cm$^2$ for RASS ($0.5-2.0$ keV) and eRASS-8 ($0.2-2.3$ keV, \citealt{Predehl2020}).} than ROSAT \citep{Predehl2020}. It will be able to make a complete detection of clusters with $M>2\times10^{14}~{\rm M}_{\odot}$, detect $>100,000$ galaxy clusters, and thus allow us to set tighter constraints on dark energy (e.g., \citealt{Merloni2012,Merloni2020,Pillepich2012,Pillepich2018}). First results indeed look very promising (e.g., \citealt{Ghirardini2021,Reiprich2021}). However, it will take time until the first full sky eROSITA cluster catalogs are published.

In \citet{Xu2018} (hereafter Paper I), we applied an X-ray, wavelet-based, source detection algorithm to the RASS data, and found many known as well as many new candidates of galaxy groups and clusters, which were not included in any previous X-ray or SZ (detected by the Sunyaev-Zel$'$dovich effect in micro-wave band) cluster catalogues. Out of those, we showed a pilot sample of $13$ very extended groups, whose surface brightness distributions are flatter than expected and, some of them have X-ray fluxes above the nominal flux-limits of previous RASS cluster catalogues. We argued that such galaxy groups were missed in previous RASS surveys due to their flat surface brightness distribution, which was not detected by the sliding-cell algorithm.

In this paper, we release our final galaxy cluster catalogue, the RASS-based X-ray selected extended Galaxy Clusters Catalog (RXGCC), and discuss the cluster properties (e.g., position and redshift, angular size, X-ray luminosity, mass, etc.). The paper is structured as follows. 
Section~\ref{sec:method_data} describes the data and methodology briefly. Section~\ref{sec:result} presents the RXGCC catalog, and 
the distribution of parameters. 
In Section~\ref{sec:discussion}, we discuss the very extended clusters and false detections in our work, and compare the RXGCC catalog with other ICM-detected cluster catalogs and a general X-ray source catalog. Section~\ref{sec:conclusion} shows the conclusion of this project. 

Additional information is added in the appendix. We discuss the difference between the method applied here and that of Paper I in Sec.~\ref{sec:diff_paper1}. 
In Sec.~\ref{sec:contam}, we estimate the contamination ratio for detections with simulations. In Sec. \ref{subsec:det_undet_mcplab}, we show the comparison of detected and undetected MCXC, PSZ2, Abell clusters. In Sec.~\ref{sec:reflex_noras}, we compare detected and undetected REFLEX and NORAS clusters. In Sec.~\ref{sec:examples}, we show the gallery of two clusters as example. In Sec.~\ref{sec:z_cla_change}, we list clusters whose redshift or classification is changed using the visual check.

Throughout the paper, we assume the cosmological parameters as H$_0 = 70$~km~s$^{-1}$~Mpc$^{-1}$, $\Omega_{\rm M} = 0.3$, 
and $\Omega_{\Lambda} = 0.7$. 

\section{Data and methodology}
\label{sec:method_data}

In this section, we present a brief description of the data and the methodology to detect, classify, and characterize sources. Further details are provided in Paper I, and method modifications are discussed in Sec.~\ref{sec:diff_paper1}.

\subsection{Data}
\label{subsec:data}

We use the RASS images in the $0.5-2.0$~keV energy band. These images come from the third processing of the RASS data (RASS-3\footnote{https://heasarc.gsfc.nasa.gov/FTP/rosat/data/pspc/\\processed$\_$data/rass/release/}). 
We exclude the regions within galactic latitude $|b|<20^\circ$,
the Virgo cluster, the Large and Small Magellanic Clouds (see Tab.~$2$ in \citealt{Reiprich2002}). The total data coverage in this work is $26,721.8$ deg$^2$ ($8.13994$~sr), about two-thirds of the sky. 

\subsection{Source detection}
\label{subsec:detection}

Our source detection and characterization method consists of three main steps. First, we apply a multi-resolution wavelet filtering procedure \citep[][]{Pacaud2006,Faccioli2018} to remove the Poisson noise in the RASS images. 
Secondly, we use the \textsc{SExtractor} software \citep{Bertin1996} on the filtered images. 
Finally, a maximum-likelihood fitting algorithm models each detected source. This last step makes use of C-statistics \citep{Cash1979} and fits two models to each source: a point source model, given by the point-spread function (PSF) of the PSPC detector, and a cluster model, described by a $\beta$-model with $\beta=2/3$ \citep{cavaliere1976}. The $\beta$-model describes the surface-brightness of galaxy clusters with a spherically symmetric model, $S_{\rm x}(r) \propto [1+(r/r_{\rm c})^2]^{-3\beta+0.5}$, where $r_{\rm c}$ is the core radius of the cluster, and $\beta$ describes the slope of the brightness profile. The value of $\beta$ is $2/3$ for a typical cluster. The smaller $\beta$, the flatter the profile. 

The maximum-likelihood fitting algorithm outputs several parameters for each source. Using extensive Monte Carlo simulations (see Paper I for further details), we
calibrate and find criteria to classify the detections. The simulations include a realistic sky and particle background, a population of 
AGN, and 
galaxy clusters with different profile,
size, 
and flux. We have chosen the {\it detection likelihood} (i.e. the significance of each detection compared with a background distribution), the {\it extension likelihood} (i.e. the significance of the source extension), and the {\it extent} (i.e. the core radius of the cluster model), as best parameters to classify our detections. It is expected the false detection rate is $0.0024$/deg$^2$ with the thresholds of {\it detection likelihood}~$>20$, {\it extension likelihood}~$>25$ and {\it extent}~$>0.67$~arcmin. 

\subsection{Classification of detections}
\label{subsec:classification}

To classify the detections, we consider not only the spatial and redshift distribution of galaxies, but also the information from previously identified clusters. As described in this subsection, we collect galaxy redshifts and identify clusters, to estimate the redshift and classify our detections based on the classification criteria. In the last step, we visually inspect images in multiple bands to confirm our redshift estimation and classification.

\subsubsection{Redshift estimation}
\label{subsec:z}

To estimate redshifts of the detections, we gather all spectroscopic and photometric redshifts of galaxies located within $15$~arcmin surrounding our detections, with $0<z<0.4$, from the Sloan Digital Sky Survey (SDSS) DR16\footnote{http://www.sdss.org}, Galaxy And Mass 
Assembly DR1\footnote{http://www.gama-survey.org} (GAMA), the Two Micron All 
Sky Survey Photometric Redshift Catalog (2MPZ catalog, \citealt{Bilicki2014}) 
and the NASA Extragalactic Database\footnote{https://ned.ipac.caltech.edu} (NED).  
We consider galaxies as the same object if the offset $<3"$ and $\Delta z<0.01$, and only reserve the information of one galaxy 
in the following priority order: SDSS, GAMA, 2MPZ, NED. The spectroscopic redshift in surveys has a higher priority than the photometric redshift.

This way, we obtain the distribution of galaxy redshifts, 
both spectroscopic and photometric ones, for each detection. 
The galaxy redshifts are binned with the width of $\Delta z=0.01$. We iteratively 
fit a Gaussian function with $3\sigma$ clipping to the two highest peaks in 
the distribution. Each peak should contain $\ge 3$ redshifts. The best fit value is taken as  the candidate redshift. Note that the second peak is also taken as 
an alternative option of the candidate redshift, in case of the included galaxy number is no less than 20$\%$ of the 
highest peak. The redshift difference of the two candidate values is
set as $\Delta z > 0.02$ to avoid the overlap.
Furthermore, the $1\sigma$ redshift 
dispersion is estimated, including the intrinsic cluster velocity dispersion and the uncertainty of redshift estimation. The redshift error is set as $\sigma_z>0.005$.

In addition, we perform a visual inspection (see Sec.~\ref{subsec:visualinspection}) 
to confirm the redshift of detections.
If there are not enough galaxy redshifts measured,
the final redshift is obtained from the collected redshift information 
of known galaxy clusters, if available. ICM-detected cluster information has 
a higher priority than optical cluster information.
All this information, if applicable, is added to the 
'$z$,src' column of Tab.~\ref{tab:candi}.

\subsubsection{Collection of identified clusters}
\label{subsec:cluster-literature}

We perform an extensive cross-matching of our detections with other
publicly available cluster catalogues to determine how many 
detections have been identified as clusters or groups previously.
We collect the information of galaxy clusters from the literature, 
and list them in Tab.~\ref{table:catalogs}.
As mentioned in Sec.~\ref{sec:introduction}, the MCXC catalog is a meta-catalog that includes a number of individual catalogs, such as the REFLEX and NORAS catalogs.
In addition, we take into account systems from the NED database, 
which are classified as \texttt{cluster of galaxies, group of galaxies, 
galaxy pair, galaxy triple}, or \texttt{group of QSOs}. We discard these 
NED systems with the name beginning with $WHL$, $MCXC$ or $PSZ2$, 
to avoid the repetition with catalogs from the literature.

Given that the Abell and ICM-detected clusters have a limited accurate redshift, we also cross-match detections with them only by projected distance without considering their redshifts, and list the cross-matching result in the last column of Tab.~\ref{tab:candi} as a reference. This information is not used to classify our detections, which is discussion in Sec.~\ref{subsec:class_criteria}.

Besides the cluster catalogs, we also list ROSAT X-ray point-source catalogs in the last part of Tab.~\ref{table:catalogs},
see Sec.~\ref{X ray sources} for details.

\begin{table*}[t]
\caption{\footnotesize{Overview of cluster catalogs and X-ray point-source catalogs.}}
    \begin{threeparttable}
    \centering
    \tiny
        \begin{tabular}{c c r r c }
        \hline
        \hline
Catalogs & Reference  & N$_{\rm gc}$ & N$_{\rm{gc(with}~z)}$ & Survey \\
        \hline
         X-ray catalogs& \citealt{Finoguenov2020}    & $10\,382$ & $10\,382$ &ROSAT\\
                       & \citealt{Takey2011,Takey2013,Takey2014,Takey2016} & $904$ & $904$ & {\it XMM-Newton}, SDSS \\
                       & \citealt{Clerc2020,Kirkpatrick2021}    & $2\,740$ & $2\,740$ & ROSAT\\
                       & \citealt{Clerc2016}    & $240$ & $240$ & ROSAT, XMM-Newton \\
                       & \citealt{Clerc2012}    & $422$ & $176^*$ &XMM-Newton XXL \\
                       &\citealt{Pacaud2016}    &    $100$ &    $100$& {\it XMM-Newton} XXL \\   
                       &\citealt{Liu2015}       &    $263$ &     $0$ & Swift\\
                       &\citealt{Mehrtens2012}  &    $503$ &    $463$ & {\it XMM-Newton}  \\
                       &\citealt{Piffaretti2011}& $1\,743$ & $1\,743$ &ROSAT\\
                       &\citealt{Ledlow2003}    &    $579$ &    $579$ & ROSAT \\
                       &\citealt{Ebeling1996}   &    $283$ &    $278$ & ROSAT \\
        \hline
        Microwave catalogs   &\citealt{Tarrio2019}      & $2\,323$ &$1\,459$ & {\it Planck}, RASS\\
                             &\citealt{Tarrio2018}      & $225$ & $171$ & {\it Planck}, RASS\\
                             &\citealt{Hilton2018}      & $182$ & $182$ &ACT \\
                             &\citealt{Planck2016a}     & $1\,653$ & $1\,094$&{\it Planck}  \\
                             &\citealt{Bleem2015}       & $677$ & $649$ & SPT \\ 
                             &\citealt{Hasselfield2013} & $91$ & $91$ & ACT \\
                             &\citealt{Marriage2011}  & $23$ & $23$ & ACT   \\
        \hline
Optical $\&$ Infrared catalogs &\citealt{Oguri2018}    & $1\,921$  & $1\,921$  & HSC \\
                            &\citealt{Oguri2014}    & $71\,743$ & $71\,743$ & SDSS \\
                            &\citealt{Rykoff2016}   & $26\,898$ & $26\,898$ & SDSS \\
                            &\citealt{Rykoff2014}   & $25\,325$ & $25\,325$ & SDSS \\
                            &\citealt{Wen2018}      & $47\,600$ & $47\,600$ & 2MASS, WISE, SuperCOSMOS \\
                            &\citealt{Wen2015}      & $25\,419$ & $25\,419$ & SDSS \\ 
                            &\citealt{Wen2012}      &$132\,684$ & $132\,684$& SDSS \\ 
                            &\citealt{Abell1958,Abell1989} & $5524$ & $875$ & -  \\
        \hline
             X-ray point-source catalogs &\citealt{Voges1999} & & & ROSAT\\
               &\citealt{Voges2000} & & & ROSAT\\ 
                    &\citealt{Boller2016} & & & ROSAT\\ 
        \hline
        \hline
        \end{tabular}
        \begin{tablenotes}
        \item[$^*$] For the reliability, we take into account clusters with redshift status as "confirmed" (spectroscopic) or "photometric", and discard those with redshift status as "provisional" or "tentative".
        \end{tablenotes}
        \end{threeparttable}
\label{table:catalogs}
\end{table*}

\subsubsection{Classification Criteria}
\label{subsec:class_criteria}

We cross-match our detections with previously identified
clusters described in Sec.~\ref{subsec:cluster-literature}, 
using a matching radius of $15'$ and $0.5$~Mpc, and $\Delta z < 0.01$. 
If all criteria are met for two clusters, we consider them as the same one. 
With these criteria and visual inspection, we classify our 
detections into four classes:
\begin{itemize}
	\item $Gold$, if the detection is not cross-identified 
	with any known optical, infrared, X-ray or SZ cluster, and 
	performs as a clear galaxy over-density in optical and 
	infrared images and/or has a clear peak in the
	distribution of galaxy redshifts.
	\item $Silver$, if the detection has been cross-identified 
	with a known optical or infrared cluster, but not cross-identified 
	with any X-ray or SZ cluster.
	\item $Bronze$, if the detection has been cross-identified 
	with known X-ray or SZ cluster.
	\item $False~detections$, if the detection seems a spurious detection,
	or if it is artificially created by multiple high-redshift clusters
	that are nearby located in projection.
\end{itemize} 

\subsubsection{Visual inspection}
\label{subsec:visualinspection}

We create an extensive multi-wavelength gallery for each of 
our detections.
For each source,
we checked the RASS photon image, RASS exposure map, wavelet-reconstructed 
image, as well as the galactic neutral hydrogen column density image from 
the HI4PI survey \citep{HI4PI2016}.
These images, combined with galaxy redshifts described in
Sec.~\ref{subsec:z}, help us to find out detections artificially created 
by the large variation of exposure time, or a low hydrogen 
column density along the line-of-sight.
Furthermore, when available, we inspected image
from the $Planck$\footnote{http://www.esa.int/Planck} survey (refer \citealt{Erler2018} for the corresponding image extraction method), infrared RGB image from 2MASS\footnote{https://www.ipac.caltech.edu/2mass},
optical image from the POSS2/UKSTU Red 
survey\footnote{http://archive.eso.org/dss/dss}, and optical RGB image
(combination of $i$, $r$ and $g$ bands) from VST 
ATLAS\footnote{http://astro.dur.ac.uk/Cosmology/vstatlas}, 
Pan-STARRS1\footnote{https://panstarrs.stsci.edu}, DES 
DR1\footnote{https://www.darkenergysurvey.org}, and SDSS DR12. 

The detections of M101 (RA: $210.828^\circ$, DEC: $54.313^\circ$) 
and M51 (RA: $202.495^\circ$, DEC: $47.211^\circ$) are
confirmed as galaxies using the visual inspection, thus
removed from the final catalog. 

In addition, visual inspection is performed to confirm, and modify if needed, the redshift and
classification of detections. 
The modifications 
are listed in Tab.~\ref{tab:app_corz}.
\begin{itemize}
\item Whether the spatial distribution of member galaxies
obeys the cluster morphology, ($yes$ for the existence of a cluster),
\item Whether a bright elliptical galaxy is found at the central region, ($yes$ for the existence of cluster), 
\item Whether the X-ray emission mainly comes from a star, star cluster, galaxy, or AGN, ($yes$ for false detection).
\end{itemize}

\subsection{Characteristics of cluster candidates}
After the source detection and classification, we  estimate physical parameters, such as the size, flux, luminosity, and mass, to characterize cluster candidates. Besides that, we also estimate their $\beta$-value to show the compactness.

\subsubsection{Estimation of physical parameters}
\label{subsec:para_estimation}

We estimate X-ray observables using the growth curve analysis method \citep{Bohringer2000,Bohringer2001}, which is briefly described in this subsection. 
Firstly, the growth curve, i.e. the cumulative count-rate of photons as a function of radius, is constructed for each cluster in the [$0.5-2.0$]~keV energy band.
The regions within the radius of $20'$, $40'$, and $60'$ are taken as the source area, respectively. In some complex cases, other radius is taken for the source area, such as $30'$, $50'$, or $80'$.
A larger annulus with the width of $20'$ is used to estimate the local background.
Contamination in the background and source area, from projected foreground or background sources, is corrected using a de-blending procedure, which excludes angular sectors with count rate different from the median value by $>2.3\sigma$. 

After the background is subtracted, we bin the growth-curve with the width of $0.5'$. The growth curve is supposed to be flat at large radius, where the contribution of cluster is negligible.
We define the significant radius, $R_{\rm sig}$, as the radius whose 
$1\sigma$ uncertainty interval encompasses all count rates integrated in larger apertures.
We take the growth curve beyond the significant radius as the plateau, and take the 
average plateau count rate as the significant count rate, $CR_{\rm sig}$. 
For each candidate, the source aperture with the most steady growth curve is used for following analysis. 

With the estimation of $R_{\rm sig}$ and $CR_{\rm sig}$, we calculate the main
parameters with the scaling relations from \citet{Reichert2011}, the Eq.~$23$-$26$ 
wherein show the power-law relations between  $L_{\rm x}$, $T_{\rm x}$, and $M$.
Firstly, we assume $R_{500}=R_{\rm sig}$, and obtain the total mass 
within $R_{500}$ ($M_{500}$), and further the bolometric X-ray 
luminosity ($L_{\rm x}$) and temperature ($T_{\rm x}$), using scaling relations. 
With the APEC model,
we obtain the luminosity within $R_{500}$ in [$0.1-2.4$] keV band ($L_{500}$), as well as the flux in the 
same band ($F_{500}$). The total count rate inside $R_{500}$ ($CR_{500}$) is further estimated
with PIMMS \citep{Mukai1993},
taking into account the instrument response of ROSAT
PSPC, and the galactic absorption. 
Using the typical $\beta$-profile with $\beta=2/3$, we take the value of $CR_{\rm 500}$ as an estimation of $CR_{\rm sig}$. The goal is that the estimated $CR_{\rm sig,est}$ is equal to the observed $CR_{\rm sig,obs}$. Therefore, the steps described above are iteratively repeated until $CR_{\rm sig,est}$ = $CR_{\rm sig,obs}$.
This way, we obtain the value of $CR_{500}$ and $R_{500}$, as well as 
corresponding parameters, such as $M_{500}$, $L_{500}$, $T_{\rm x}$, and $F_{500}$. 

\subsubsection{Estimation of $\beta$ value}
\label{sec:get_beta}

To better understand the clusters in our sample, we characterise their X-ray emission profile. As mentioned in Sec.~\ref{subsec:detection},  the $\beta$ value describes the flattening of the source. And the $\beta$ parameter is highly correlated with the core radius.

Using the Markov Chain Monte Carlo (MCMC) method, we estimate the value and uncertainty of the $\beta$ and the core radius, by fitting the growth curve with a $\beta$-model convolved with the PSF. For the simplicity, we take a gaussian function with the FWHM of $45$~arcsec as the PSF of the PSPC detector. In the MCMC fitting\footnote{EMCEE package: https://emcee.readthedocs.io/en/stable/}, we use $50$ chains in total, $10\,000$ steps each. The free parameters include the $\beta$ value, the core radius, and the normalization. We initialize chains in a tiny range around the maximum likelihood result, and discard first $5\,000$ steps of each chain to avoid the effect from the starting point. Then, a uniform prior is assumed for each parameter, whose threshold is set as respectively $0.50$ and $1.00$ for $\beta$ value, $0.01$ and $50.0$~arcmin for the core radius, $0.001$ and $5$ for the normalization. The lower limit of the $\beta$ value comes from the convergence requirement of the integrated count rate.

\section{Results}
\label{sec:result}

\begin{figure}[t]
\centering
\includegraphics[width=0.48\textwidth]{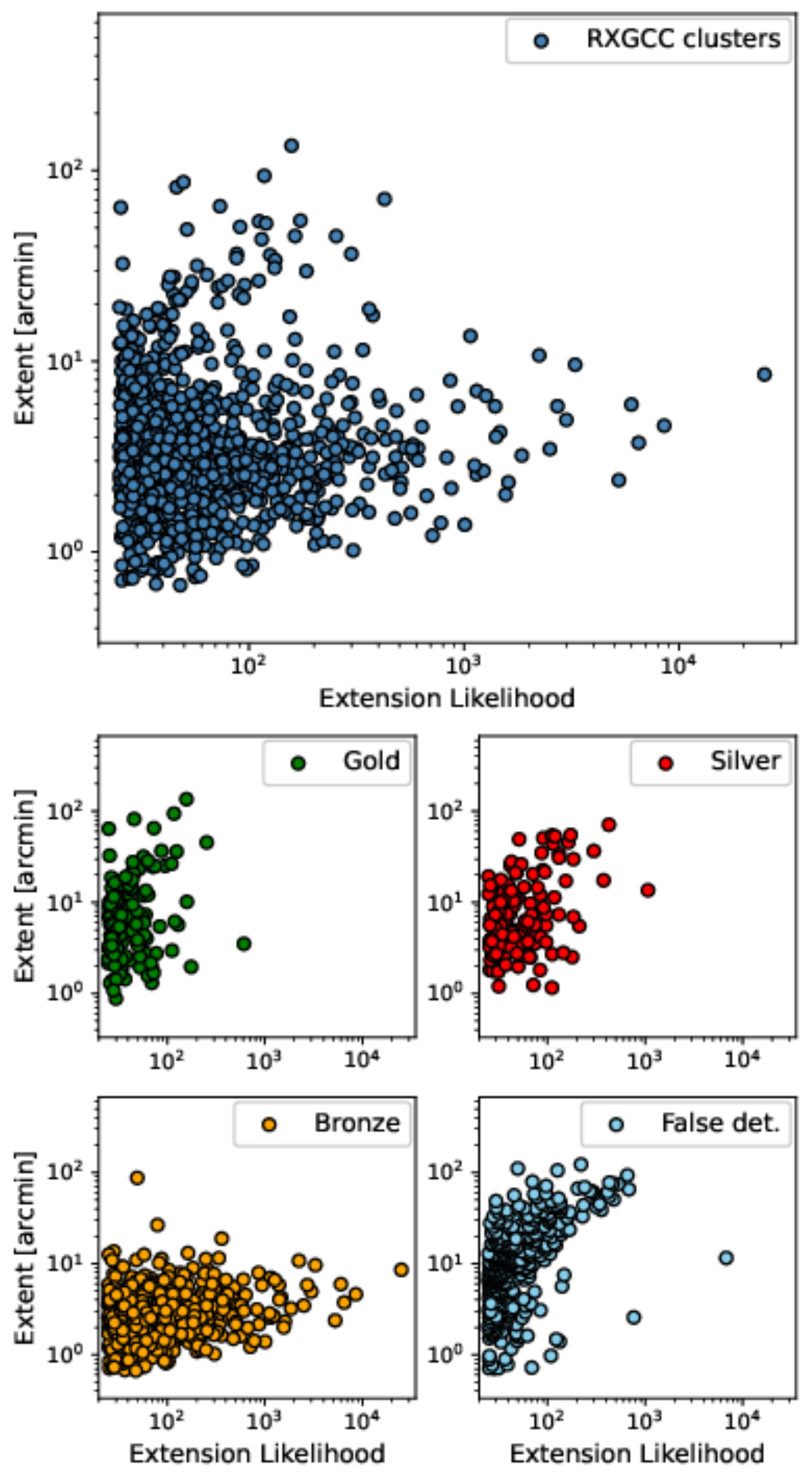} 
\caption{\footnotesize
{$Extent$ - $extension~likelihood$ plane that 
defined our criteria 
to select extended X-ray sources (see Paper I for further 
details). In the top panel, we show the RXGCC clusters, i.e., $Gold$+$Silver$+$Bronze$.
The following top-left, top-right, and bottom-left panels show sub-samples of RXGCC: $Gold$ - new identified clusters, $Silver$ - previous
optical/infrared-identified clusters, $Bronze$ - previous ICM-identified clusters. The bottom-right panel shows the $False~detection$ - spurious detections.}}
\label{fig:ext_extml_cla}
\end{figure}

We obtain $1\,308$ detections with the RASS data in the high galactic latitude area. 
Following the source detection, classification and characterization methods described in Sec.~\ref{sec:method_data}, we obtain $641$ $Bronze$ clusters, $154$ $Silver$ clusters, and $149$ $Gold$ clusters. These $944$ cluster candidates are compiled into the final catalog, named as the RASS-based X-ray 
selected Galaxy Cluster Catalog (RXGCC).
A detailed discussion about false detections is provided in 
Sec.~\ref{subsec:false_detections}. The distribution of each class in 
the {\it extent} - {\it extension likelihood} parameter space 
(see Sec.~\ref{subsec:detection}) are shown in Fig.~\ref{fig:ext_extml_cla}, 
and the source number and median redshift are summarized in
Tab.~\ref{tab:n_class}. 

\begin{table}[ht]
 \begin{center}
  \caption{\footnotesize{The detections in classes.}}
   \label{tab:n_class}
    \begin{tabular}{l c c}
        \hline
        \hline
    Class	   &	Number  & Median	redshift\\ 
        \hline
    $Gold$	    &  $149$	& $0.0914$ \\
    $Silver$	&  $154$    & $0.0584$ \\
    $Bronze$	&  $641$	& $0.0780$ \\
    $False~detections$& $364$	& - \\
        \hline
        \hline
        \end{tabular}
 \end{center}
\end{table}

\begin{figure}[t]
\centering
\includegraphics[width=0.45\textwidth]{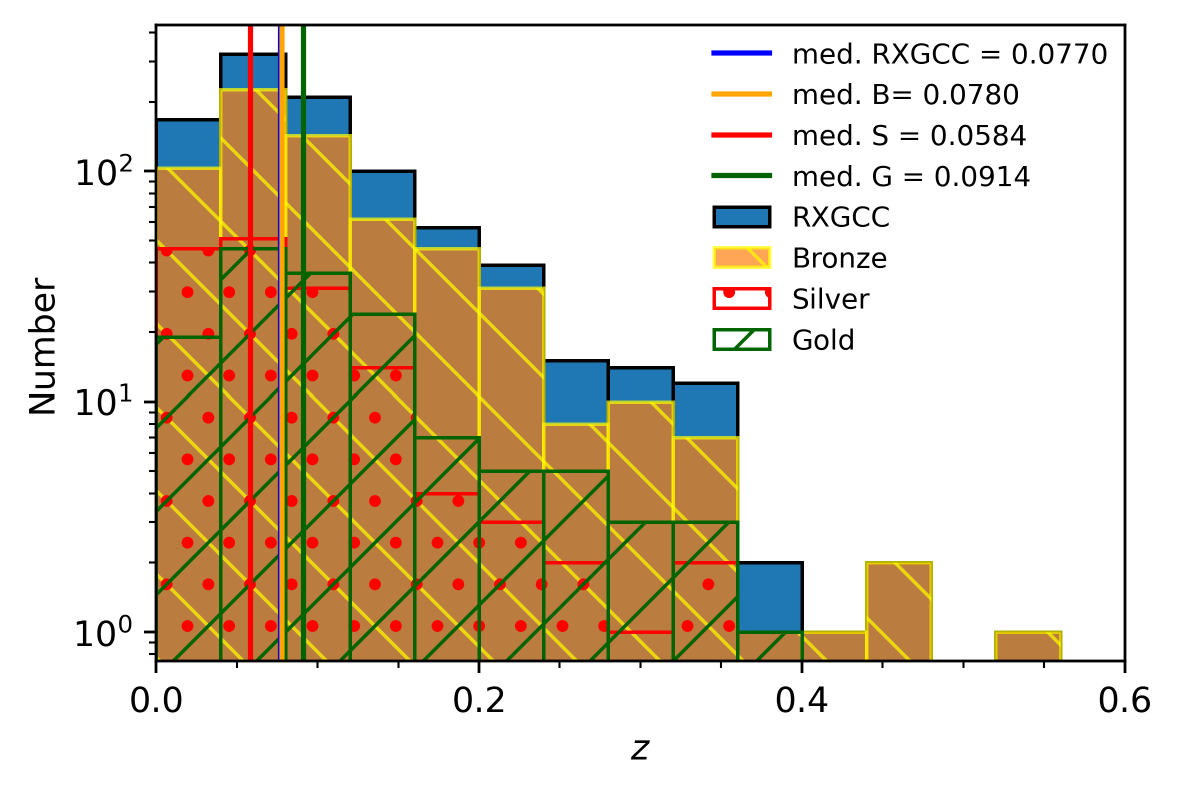} 
\caption{\footnotesize
{Redshift distribution for different classes of our 
cluster candidates. The overlaid vertical line indicates 
the median redshift of different classes. For the simplicity, we 
use 'G', 'S', and 'B' for short of the $Gold$, $Silver$, 
and $Bronze$ sample.}}
\label{fig:histz_cla}
\end{figure}

The redshift distribution of RXGCC clusters is shown 
in Fig.~\ref{fig:histz_cla}. It shows that the newly-identified 
clusters ($Gold$) tend to have a little higher redshift than the clusters previous identified as optical/infrared clusters ($Silver$),
while the previously ICM-identified 
clusters ($Bronze$) have a consistent redshift distribution with the whole RXGCC sample, whose median redshift is $0.0770$. 
The redshift histogram demonstrates the detection efficiency 
decreases greatly when the redshift $z>0.12$. 

The highest redshift of the sample is $0.5551$, and there are 
four RXGCC clusters with $z>0.4$ (RXGCC~$229$,  RXGCC~$237$,  RXGCC~$306$, RXGCC~$590$).  
These redshifts are obtained from 
previous ICM-detected clusters, instead of galaxy redshifts. 
A careful visual check of X-ray and micro-wave images, 
the spatial distribution and distance between these four 
detections and previous-identified clusters, 
makes the evidence strong enough to say the detected X-ray 
emission indeed comes from the
previously-identified clusters.
For these four candidates, the redshift limit 
of cross-matched clusters is set as $z<0.6$, instead of $z<0.4$, and the offset threshold is $<1.5$~Mpc, instead of $<0.5$~Mpc. 

In addition, three clusters, RXGCC~$406$, RXGCC~$639$, and RXGCC~$646$, are 
covered with deep optical observation. In the redshift estimation, 
galaxy redshifts are constrained as $<0.16$, instead of $<0.4$.  

For the cross-matching of RXGCC with previous ICM-detected clusters, 
there are several exceptions. 
RXGCC~$93$, RXGCC~$239$, RXGCC~$612$, RXGCC~$688$ have ICM-identified clusters in 
$<15'$, but the position and redshift distribution of 
galaxies indicate clearly that the previous ICM-detect clusters 
are not what we 
detected. Thus, we remove them out of $Bronze$ and classify 
them as $Gold$ or $Silver$ further.  

\subsection{RXGCC catalog}

\begin{figure}[t]
\centering
\includegraphics[width=0.48\textwidth]{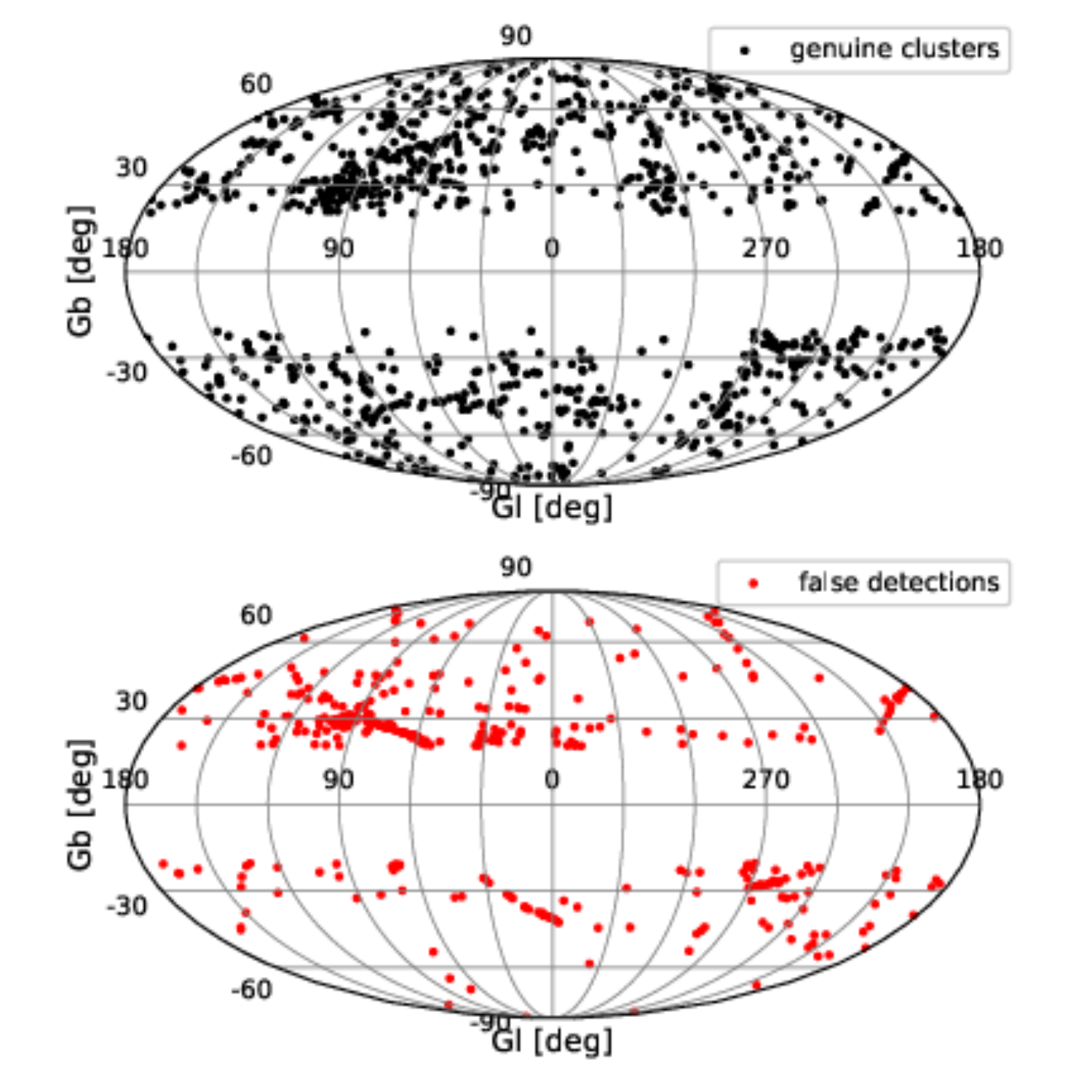} 
\caption{\footnotesize
{$Top$: The all-sky distribution of RXGCC clusters. 
$Bottom$: The all-sky distribution of false detections.}} 
\label{fig:allsky}
\end{figure}

The RXGCC catalog comprises $944$ clusters, including $641$ $Bronze$, $154$ $Silver$, and $149$ $Gold$ clusters, whose spatial 
distribution is shown in the top
panel of Fig.~\ref{fig:allsky}. 
The full table is sorted with Right Ascension and is 
provided in FITS format as supplement material (\url{https://github.com/wwxu/rxgcc.github.io/blob/master/table_rxgcc.fits}).
The first $10$ rows of the RXGCC catalog are shown in Tab.~\ref{tab:candi} and Tab.~\ref{tab:candi2} for its 
form and content.
In addition, we also publish the image gallery for each RXGCC source. For example, the gallery for RXGCC.~$1$ is shown 
in the following page, \url{https://github.com/wwxu/rxgcc.github.io/blob/master/script/1.md}.

In Tab.~\ref{tab:candi}, the first five columns are the 
name, position, $extent$ and $extension~likelihood$ values. 
The RXGCC cluster coordinates are obtained from the 
maximum-likelihood fitting (see Sec.~\ref{subsec:detection}). 
The 6th and 7th 
columns are the estiamtion of redshift and its error
(see Sec.~\ref{subsec:z}). The 8th column shows the source of the 
cluster redshift, referring the table note for the detail. The 9th column
shows the classification, described in Sec.~\ref{subsec:class_criteria}.
The following two columns contain information on the cross-matching with
previously ICM-identified clusters, and clusters previously identified in optical or infrared band. 
These clusters are all offset from our candidates 
$<0.5~$Mpc and $<15'$, and the redshift difference $\Delta z < 0.01$. 
The last column lists the Abell clusters or ICM-detected 
clusters within $0.5~$Mpc and $15'$, regardless of their redshift value. 

\begin{sideways}
    \begin{threeparttable}[t]
\centering
  \caption{\footnotesize{The first part of the first $10$ entries in the RXGCC catalog.}}
   \label{tab:candi}
    \tiny
    \begin{tabular}{c c c c c c c c c c c c }
        \hline
        \hline
RXGCC   &RA         &Dec    &Extent &Ext,ml&$z$&$z$,err& $z$,src\tnote{**} &C\tnote{\dag}&GC(XSZ,$\Delta z<0.01$)\tnote{+} &GC(OPT,$\Delta z<$0.01)\tnote{-}&GC\tnote{$\star$}\\
&J2000 [$^\circ$]  &J2000 [$^\circ$] &[arcmin] & & & & & & & \\
(1) & (2) & (3) & (4) & (5) & (6) & (7) & (8) & (9) & (10) & (11) & (12) \\
        \hline
~1  &0.026  & ~~~8.279  &	2.13  &	~56.61  &	0.0364  &	0.005  &	z1, z$\_$xsz   & B  & MCXC                             & N                        & C, F20, MCXC, N, SPI, W \\
~2  &0.641  & ~~~8.451  &	6.41  &	~42.14  &	0.0961  &	0.005  &	z1, z$\_$xsz   & B  & F20, SPI	                       & A, W                   & A, C, F20, N, SPI, W\\
~3  &0.797  &$-$35.946  &	3.04  &	128.14  &	0.0493  &	0.005  &	z1, z$\_$xsz   & B  & MCXC, Tar, XB                 & A, N                     & A, MCXC, N, Tar, W, XB\\
~4  &1.339  & ~~16.223  &	3.36  &	~85.08  &	0.1154  &	0.005  &	z1, z$\_$xsz   & B  & F20, MCXC                        & C, N, RM, W       & A, C, F20, MCXC, N, W\\
~5  &1.497  &$-$34.702  &	2.12  &	~55.62  &	0.1145  &	0.005  &	z1, z$\_$xsz   & B  & MCXC, PSZ2, Tar, XB           & A, W                   & A, MCXC, N, PSZ2, Tar, W, XB\\
~6  &1.585  & ~~10.866  &	1.73  &	~28.41  &	0.1666  &	0.005  &	z1, z$\_$xsz   & B  & F20, MCXC, PSZ2, SPI, Tar        & C, N, RM, W, Zw   & C, F20, MCXC, N, PSZ2, SPI, Tar, W\\
~7  &2.556  & ~~~2.093  &	8.87  &	~25.46  &	0.1577  &	0.005  &	z1,	           & G  & -                                & -                        & C, N, W\\
~8  &2.700  &$-$81.160  &	6.66  &	~26.49  &	0.1246  &	0.005  &	z1, z$\_$opt   & S  & -	                               & W                      & W\\
~9  &2.823  &$-$28.848  &	2.07  &	136.69  &	0.0616  &	0.005  &	z1, z$\_$xsz   & B  & MCXC, PSZ2, Tar, XB           & A, W 	              & A, MCXC, N, PSZ2, Tar, W, XB\\
10  &2.935  & ~~32.415  &	1.32  &	~61.24  &	0.1036  &	0.005  &	z1, z$\_$xsz   & B  & F20, MCXC, PSZ2, SPI, Tar, XB & A, C, RM, W	      & A, C, F20, MCXC, PSZ2, SPI, Tar, W, XB\\
        \hline
        \hline
        \end{tabular}
        \begin{tablenotes}\tiny
        \item[\dag] For the simplicity, we 
use 'G', 'S', and 'B' for short of the $Gold$, $Silver$, 
and $Bronze$ class.
        \item[**]{The source of the redshift. $'$z1$'$ means the redshift comes from the highest peak of the redshift histogram of galaxies, while $'$z2$'$ for the second highest peak. $'$z$\_$opt$'$ and $'$z$\_$xsz$'$ shows the redshift of candidate matches with or comes from (when no $'$z1$'$ nor $'$z2$'$ available) the previous identified optical clusters or ICM-detected clusters. }
        \item[+]{ 
Previously ICM-identified clusters within $0.5$~Mpc, 
$15'$ and $\Delta z<0.01$: F20 \citep{Finoguenov2020},
Tak \citep{Takey2011,Takey2013,Takey2014,Takey2016},
SPI (\citealt{Clerc2012},\citealt{Clerc2016},\citealt{Clerc2020},\citealt{Kirkpatrick2021}),
XLSSC \citep{Pacaud2016}, 
SWXCS \citep{Liu2015},
XCS \citep{Mehrtens2012},
MCXC \citep{Piffaretti2011}, 
L03 \citep{Ledlow2003}, 
XB (XBACs, \citealt{Ebeling1996}), 
Tar \citep{Tarrio2018, Tarrio2019},
H18 \citep{Hilton2018},
PSZ2 \citep{Planck2016a},
B15 \citep{Bleem2015},
H13 \citep{Hasselfield2013},
M11 \citep{Marriage2011}.}
\item[-]{ 
Clusters detected in optical or infrared band within $0.5$~Mpc, $15'$ and $\Delta z<0.01$:
C (CAMIRA, \citealt{Oguri2014, Oguri2018}), 
Zw \citep{Zwicky1968},
RM (redMaPPer, \citealt{Rykoff2014, Rykoff2016}), 
W \citep{Wen2012,Wen2015,Wen2018},
A \citep{Abell1958, Abell1989},
N for clusters from NED database.}
\item[$\star$]{The Abell clusters and ICM-detected clusters within $0.5$~Mpc and $15'$, regardless of their redshifts.}
\item[]{(This table is available in the online journal. The first ten entries 
are shown here for its form and content.) }
        \end{tablenotes}
    \end{threeparttable}
\end{sideways}

\begin{sideways}
\centering
    \begin{threeparttable}[ht]
\centering
  \caption{\footnotesize{The second part of the first $10$ entries of the RXGCC catalog. }}
   \label{tab:candi2}
    \tiny
    \begin{tabular}{c c c c c c c c c c c c}
        \hline
        \hline
   RXGCC& $R_{\rm sig}$	&$R_{\rm 500}$ & $R_{\rm 500}$ &$CR_{\rm sig}$	&$CR_{\rm 500}$	&$L_{\rm 500}$\tnote{\dag} &	$F_{\rm 500}$\tnote{\dag} &	$M_{\rm 500}$\tnote{\dag} &	$T_{\rm x}$\tnote{\dag} & Cnt$_{\rm sig}$\tnote{-} & $\beta$\\ 
   & [arcm]& [arcm] &[Mpc]&[c/s] &[c/s]&[$10^{44}$~erg/s]&[$10^{-12}$~erg/s/cm$^2$]&[$10^{14}$~M$_{\odot}$] &[keV] & \\ 
(1) & (2) & (3) & (4) & (5) & (6) & (7) & (8) & (9) & (10) & (11) & (12) \\
        \hline
 ~1 & 13.188	&	14.734 & 0.639 & 0.265 (0.037) &	0.270 (0.038) &	0.137 (0.013)&	4.457 (0.410)&	0.77 (0.04)&	1.84 (0.05)& 128	       &$0.74$    \\ 
 ~2 & 40.600	&	~9.572 & 1.023 & 0.442 (0.071) &	0.392 (0.063) &	1.677 (0.462)&	7.202 (1.982)&	3.33 (0.45)&	4.64 (0.40)& 303$^*$   &$0.50$	\\
 ~3 & 31.612	&	14.921 & 0.864 & 0.735 (0.084) &	0.676 (0.077) &	0.772 (0.067)& 13.443 (1.158)&	1.92 (0.08)&	3.25 (0.09)& 227~   &$0.52$    \\ 
 ~4 & ~9.775	&	~7.912 & 0.993 & 0.254 (0.030) &	0.246 (0.029) &	1.587 (0.096)&	4.606 (0.278)&	3.11 (0.09)&	4.47 (0.08)& 155~   & $0.91$	\\
 ~5 & 21.738	&	~8.811 & 1.098 & 0.464 (0.072) &	0.423 (0.066) &	2.859 (0.304)&	8.437 (0.898)&	4.20 (0.22)&	5.38 (0.18)& 144~   & $0.53$	\\
 ~6 & 28.156	&	~6.397 & 1.094 & 0.221 (0.056) &	0.196 (0.050) &	2.790 (0.436)&	3.641 (0.568)&	4.39 (0.33)&	5.62 (0.27)& 121~   & $0.55$	\\
 ~7 & 21.244	&	~5.677 & 0.928 & 0.116 (0.060) &	0.103 (0.053) &	1.353 (1.026)&	1.993 (1.511)&	2.65 (0.99)&	4.11 (0.97)& ~67~   & $0.89$	\\
 ~8 & 21.738	&	~7.675 & 1.029 & 0.268 (0.072) &	0.242 (0.065) &	1.794 (0.413)&	4.413 (1.016)&	3.50 (0.39)&	4.82 (0.35)& ~42$^*$& $0.88$	\\
 ~9 & 12.212	&	13.041 & 0.930 & 0.621 (0.056) &	0.627 (0.057) &	1.136 (0.056)& 12.456 (0.617)&  2.42 (0.06)&	3.77 (0.06)& 193~   & $0.76$	\\
 10 & ~9.775	&	~9.479 & 1.082 & 0.447 (0.045) &	0.445 (0.045) &	2.266 (0.096)&	8.287 (0.349)&	3.98 (0.08)&	5.19 (0.07)& 154~   & $0.78$	\\
         \hline
        \hline
        \end{tabular}
        \begin{tablenotes}\tiny
        \item[]{The X-ray observable of RXGCC clusters. The error are shown in the brackets.}
        \item[\dag]{Parameters derived with scaling relation from the count rate. The quoted uncertainty refers to the $1\sigma$ error of the count rate at significant radius, together with the uncertainty of the scaling relation.}
        \item[-]{The sum of the photon counts within the significant radius, with background photons subtracted.
        The [$^*$] symbol in this column labels out the case when the RASS tile fails to cover the whole area within the significant radius, which causes the under-estimation of the total photon number.}
        \end{tablenotes}
    \end{threeparttable}
\end{sideways}

\subsection{Parameter distributions}
\label{subsec:para}

\begin{figure*}[t]
\centering
\includegraphics[width=1.0\textwidth]{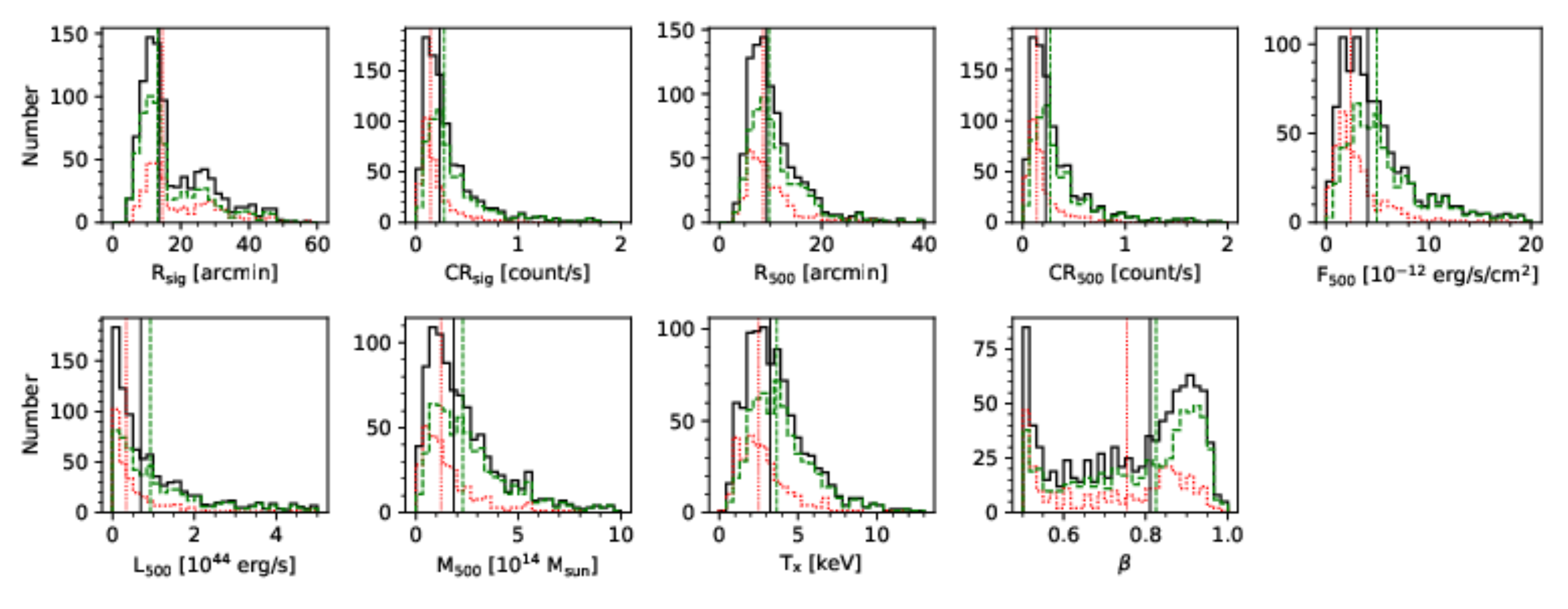} 
\caption{\footnotesize
{The distribution of parameters of the RXGCC sample, previous ICM-detected sample ($Bronze$), and new ICM-detected sample ($Gold$+$Silver$), shown with solid, dashed, and dotted histograms and lines, in sequence. Refer to Sec.~\ref{subsec:para_estimation} for details about the estimation methods. Both the $F_\mathrm{500}$ and $L_\mathrm{500}$ are the value in [$0.1-2.4$]~keV band. To show the main character, we ignore a few outliers with large parameter value in the histogram, and set manually the upper parameter limit of the axis in the plot.}}
\label{fig:para}
\end{figure*}

In this subsection, we discuss the distribution of X-ray observables of 
RXGCC clusters. 
Their distributions are shown in Fig.~$\ref{fig:para}$. The solid, dashed, dotted histograms demonstrate the parameter distribution of RXGCC sample, previous ICM-detected ($Bronze$) sample, and new ICM-detected($Gold$+$Silver$) sample, while vertical lines show the corresponding median values.
These median values are listed in Tab.~\ref{tab:para_med}.
From the median values, it is clear that the new ICM-detected clusters tend to have lower count rate, flux, luminosity, mass and temperature. Still, it is worth noting that the median flux of new ICM-detected clusters
is close to the typical flux limit of previous RASS-based cluster catalogs 
($3\times10^{-12}$~erg/s/cm$^2$). Thus, a significant fraction of new ICM-detected clusters ($Gold$+$Silver$) have flux above it.

What is more important, the new ICM-detected sample has a lower median $\beta$ value than the previous ICM-detected sample, shown in the last row of Tab.~\ref{tab:para_med}. The median $\beta$ value of clusters in these two samples is $0.76$ and $0.83$, respectively.
This is an indication for the new ICM-detected clusters to have generally flatter surface brightness profiles compared with previous ICM-detected clusters.

\begin{table}[t]
    \centering
    \caption{The median parameter value for the new ICM-detected sample ($Gold$+$Silver$), the previous ICM-detected sample ($Bronze$), and the whole RXGCC sample. Both the $F_\mathrm{500}$ and $L_\mathrm{500}$ are the value in [$0.1-2.4$]~keV band.}
    \tiny
    \begin{tabular}{l | c c c}
         \hline 
         Parameter & Med.$_{(\rm G+S)}$ & Med.$_{(\rm B)}$ & Med.$_{(\rm  RXGCC)}$ \\
         \hline 
$R_\mathrm{sig}$ [arcmin]                           &$14.650$  & $13.188$ &$13.675$\\
$R_\mathrm{500}$ [arcmin]                           & $8.534$  & $9.726$  & $9.273$ \\
$CR_\mathrm{sig}$ [c/s]                             & $0.148$  & $0.276$  & $0.234$ \\
$CR_\mathrm{500}$ [c/s]                             & $0.142$  & $0.269$  & $0.225$ \\ 
$F_\mathrm{500}$ [10$^{-12}$ erg/s/cm$^2$]          & $2.496$  & $4.955$  & $4.043$ \\
$L_\mathrm{500}$ [10$^{44}$ erg/s]                  & $0.342$  & $0.912$  & $0.688$ \\
$M_\mathrm{500}$ [10$^{14}$ M$_\odot$]              & $1.246$  & $2.270$   & $1.843$ \\
$T_\mathrm{x}$ [keV]                                & $2.514$  & $3.644$  & $3.224$ \\
$\beta$                                             & $0.76$  & $0.83$  & $0.81$ \\
         \hline
    \end{tabular}
    \label{tab:para_med}
\end{table}

\begin{figure}[t]
\centering
\includegraphics[width=0.5\textwidth]{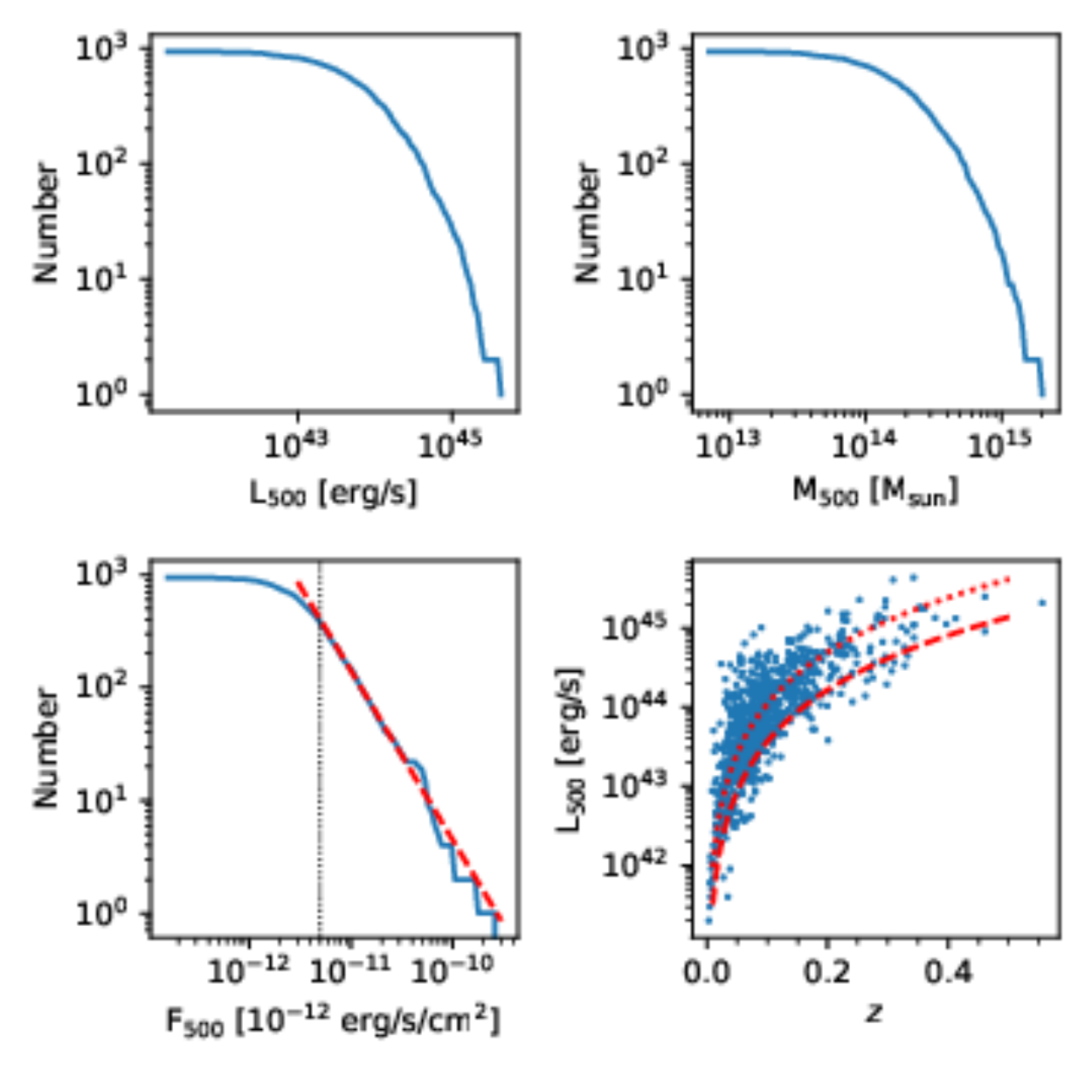} 
\caption{\footnotesize
{In top-left, top-right, and bottom-left panels, we show the relation between the value of $L_{500}$,  $M_{500}$, $F_{500}$, and the integrated cluster numbers with larger luminosity, mass, or flux.
In the bottom-left panel, the dashed line has a slope of $-1.5$, normalized to match the value with $f_{\rm x}=2\times 10^{-11}$ erg/s/cm$^2$. The dotted vertical line labels the flux of $5\times 10^{-12}$ erg/s/cm$^2$. In the bottom-right panel, it shows the relation between the redshift and $L_{500}$. The curves for the flux limits of $10^{-12}$ and $3\times 10^{-12}$ erg/s/cm$^2$ are overlapped as dotted curve and dashed curve, respectively. Both the $F_\mathrm{500}$ and $L_\mathrm{500}$ are the value in [$0.1-2.4$]~keV band.}}
\label{fig:lf_mf}
\end{figure}

In the top panels of Fig.~\ref{fig:lf_mf}, the 
measured luminosity function and mass functions are shown. Note that we have not applied any correction of selection effects here -- as a complete sample with accurate selection function would be required for this -- so a quantitative comparison to
previous catalogs, such as the REFLEX or HIFLUGCS samples, is not meaningful.
In this plot, we show the cumulative number of RXGCC clusters above a specific value of luminosity or mass, as a very rough demonstration of the luminosity and mass function. 

In bottom-left panel of Fig.~\ref{fig:lf_mf}, we show the cumulative number as a function of the X-ray flux (also called log$N$-log$S$ plot). In this plot, we overlay the dashed line with the slope of $-1.5$ predicted by a static Euclidean universe with clusters uniformly distributed, which is a reasonable assumption for low-redshift clusters for a rough completeness check. This line is normalized to match it with the measurement at $F_{500}=2\times10^{-11}$ erg/s/cm$^{-2}$.
The theoretical line matches with our detection curve 
with $F_{500}>5\times10^{-12}$ erg/s/cm$^{-2}$, shown with the vertical dotted line. 
This is a good indication that we have achieved a high completeness up 
to this flux, despite of our requirement of detecting significantly extended X-ray emission. 
A small bump at $\sim 6\times10^{-11}$~erg/s/cm$^{-2}$
is worth to be noted, which likely stems from small number statistics 
and expected large cosmic variance at low redshift; i.e., we likely 
see large-scale structure here. 

In addition, the relation between the redshift and luminosity are shown in the bottom-right panel, overlaid with the flux limit of $10^{-12}$ and $3\times 10^{-12}$ erg/s/cm$^2$ as dotted curve and dashed curve, respectively.
It is noted that there are $115$ new ICM-detected RXGCC 
clusters with flux larger than the flux limit of REFLEX 
($3\times10^{-12}$~erg~s$^{-1}$~cm$^{-2}$ in $0.1-2.4$~keV band).

\subsection{Comparison of radii, count rates, and flux}

\begin{figure}[t]
\centering
\includegraphics[width=0.5\textwidth]{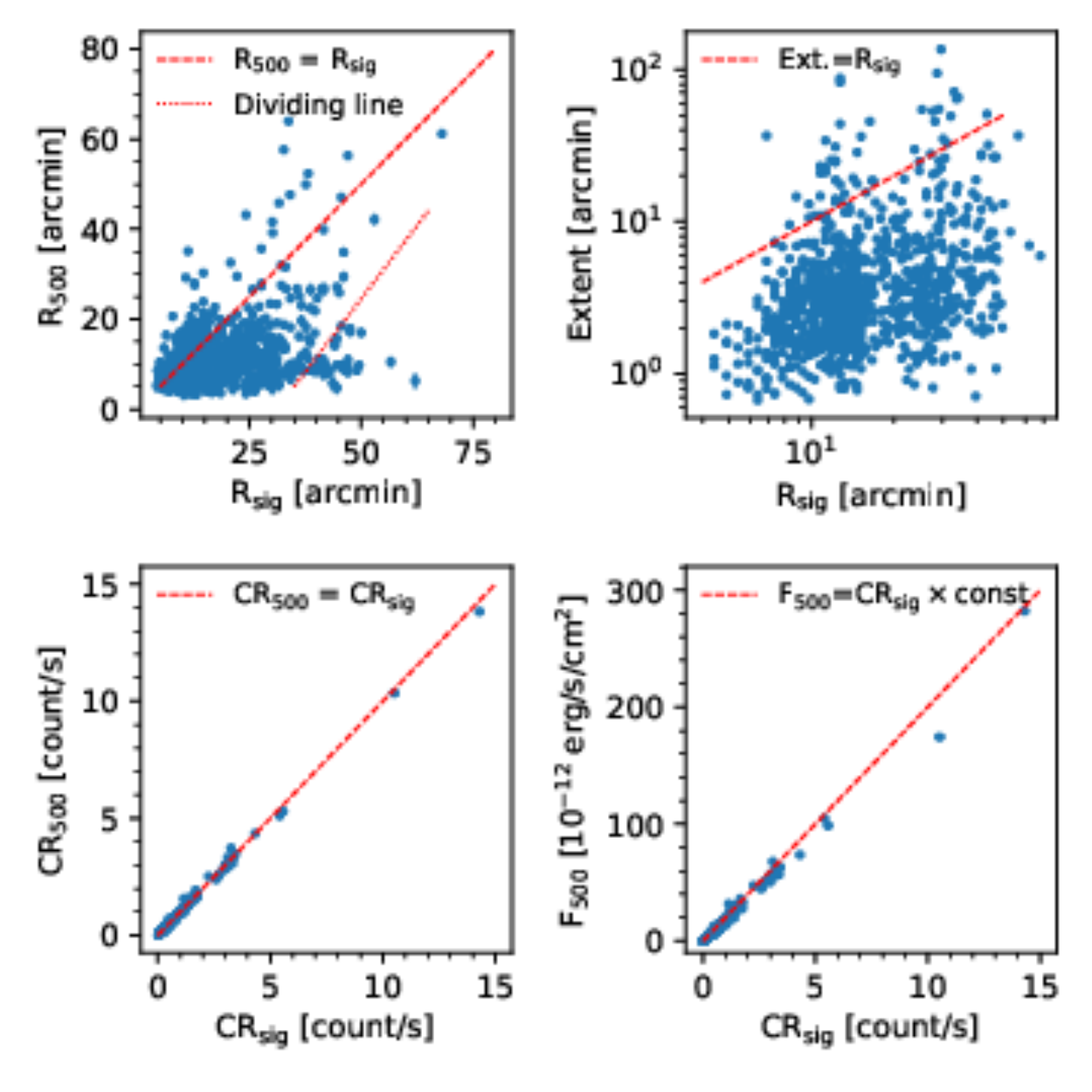} 
\caption{\footnotesize
{The relation between the $R_{500}$, $R_{\rm sig}$, $extent$, the integrated count rate within $R_{500}$, the integrated count rate within $R_{\rm sig}$, and the flux inside $R_{500}$. In the top-left, top-right, and bottom-left panels, the red dashed line labels the $1$:$1$ relation, while in the bottom-right panel it shows the relation of $F_{500}=CR_{\rm sig} \times 2 \times 10^{-11}$ (erg/s/cm$^2$)/(count/s). In the top-left panel, the red dotted line shows the dividing line to get the very extended candidates discussed in Sec.~\ref{subsec:extended_candi}.}}
\label{fig:r_cr}
\end{figure}

Firstly, we compare different radii (i.e., the significant radius, $R_{500}$, $extent$). As shown in the top panels of Fig.~$\ref{fig:r_cr}$, $R_{\rm sig}$ is compared with $R_{500}$ and $extent$, the definitions and estimation methods are described in Sec.~\ref{subsec:para_estimation} and Sec.~\ref{subsec:detection}. In the plot, it is clear that $R_{\rm sig}$ is larger than both $R_{500}$ and $extent$ in most cases.
Since $extent$ shows the core radius of the $\beta$-model, and $R_{\rm sig}$ demonstrates the region with significant X-ray emission, it is reasonable $R_{\rm sig}$ is larger in most cases. Given the low background of the ROSAT PSPC, the X-ray emission happens to spread beyond $R_{500}$. However, for some clusters we seem to detect emission out to $\sim3-5 \times R_{500}$, which is surprising, at first sight.
One possible reason is that our estimation is based on the assumption of the $\beta$-model with the typical value of $\beta=2/3$, which does not necessarily describe properly the data, especially for the very extended clusters that we aim to discover. With a much flatter profile of $\beta\ll 2/3$, the very extended cluster has an X-ray emission extending up to a much larger area. 

In addition, we compare the integrated count rate and flux within $R_{\rm sig}$ and $R_{500}$, respectively.
This is shown in the bottom panels of Fig.~$\ref{fig:r_cr}$. Compared to the top panels, these relations are much tighter. However, the conversion between the $CR_{\rm sig}$ and the $CR_{500}$ are derived by assuming a typical cluster profile with $\beta=2/3$, thus their relation is related to the relation between $R_{500}$ and $R_{\rm sig}$. The conversion between the $CR_{500}$ and $F_{500}$ is derived with the information of cluster redshift, ICM temperature, galactic absorption, and metallicity. The tight relations between the $CR_{\rm sig}$, $CR_{500}$, and $F_{500}$ indicate the robustness of our parameter estimation method. 

\begin{figure}[t]
    \centering
    \includegraphics[width=0.5\textwidth]{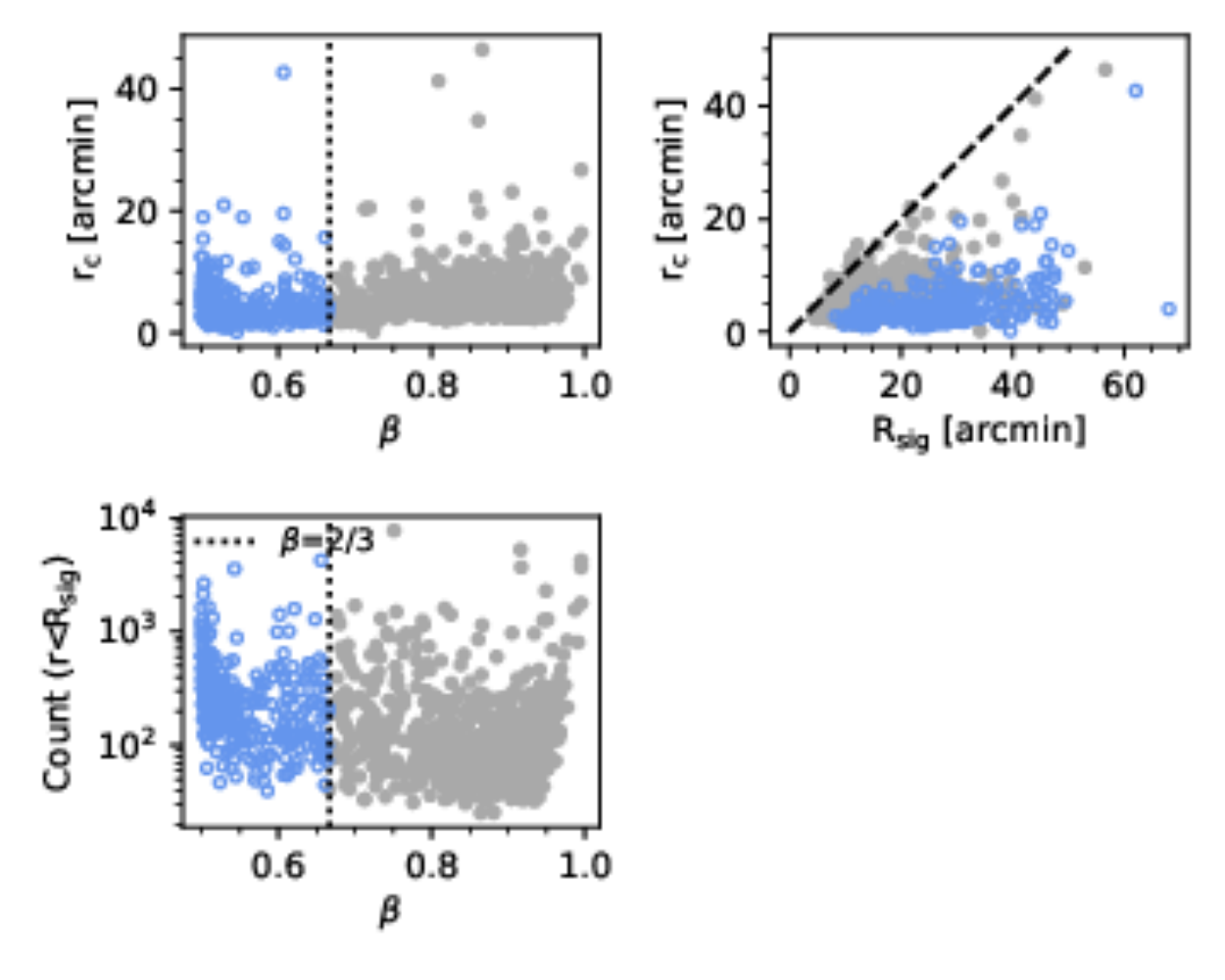} 
    \caption{The relation between the $\beta$ value, the core radius ($r_{\rm c}$), the significant radius ($R_{\rm sig}$), and the total number of photons within the significant radius. The grey dots, and blue empty circles show the 'high $\beta$' and 'low $\beta$' samples, respectively, which are divided by the line of $\beta=2/3$.}
    \label{fig:beta_hl}
\end{figure}

Furthermore, we separate the RXGCC sample into two sub-samples by the $\beta$ value, and show the relation of the $\beta$ value, the significant radius, the core radius, and the total photon count inside the significant radius in Fig.~\ref{fig:beta_hl}.
The $\beta$ value and $r_{\rm c}$ are estimated by fitting the growth curve with the $\beta$-model convolved with PSF as described in Sec.~\ref{sec:get_beta}, and $R_{\rm sig}$ is derived from the growth curve as shown in Sec.~\ref{subsec:para_estimation}. A smaller $\beta$ value corresponds to a flatter emission profile.
As shown in the bottom panel of Fig.~\ref{fig:beta_hl}, it seems there are two subsamples, thus we separate the RXGCC sample into 'high-$\beta$' and 'low-$\beta$' sample with $\beta=2/3$, and show them with different symbols. Except for the value difference of $\beta$, these two subsamples do not show much difference in the core radius, photon number within the significant radius, and the relation between the significant radius and the core radius. The significant radius is larger than the core radius for most of detections.

\begin{figure}[t]
    \centering
    \includegraphics[width=0.5\textwidth]{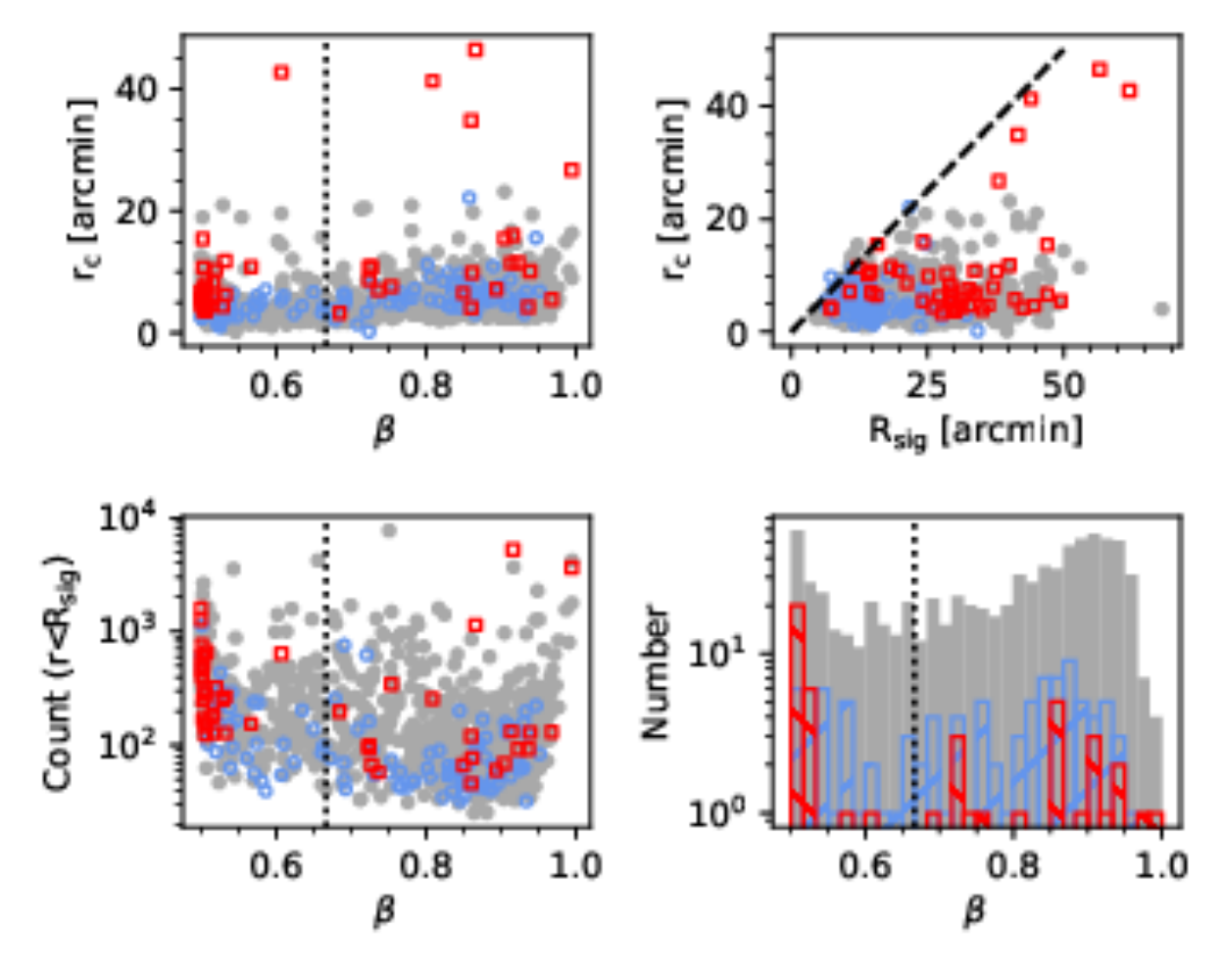} 
    \caption{Top-left, top-right, bottom-left panels are the same with Fig.~\ref{fig:beta_hl}, and only differs in symbols. In these panels, the grey dots, blue empty circles, and red empty squares represent the $Silver+Bronze$ sample, $Faint~Gold$ sample, and $Bright~Gold$ sample, respectively. The $Bright~Gold$ and $Faint~Gold$ samples are divided from the $Gold$ sample by $F_{500}=3\times10^{-12}$~erg/s/cm$^2$. Bottom-right panel shows the $\beta$ histogram of the same samples, with colors matching with other panels.}.
    \label{fig:beta_rc_rsig}
\end{figure}

Finally, we compare the same set of parameters for the $Silver+Bronze$ sample (shown with solid dots) with the $Gold$ sample (shown with empty circles and squares). As shown in Fig.~\ref{fig:beta_rc_rsig}, the later sample tends to have lower $\beta$ value, which indicates their flat profile. We further divided the $Gold$ sample into $Bright~Gold$ sample (shown with empty squares) and $Faint~Gold$ sample (shown with empty circles) by $F_{500}=3\times10^{-12}$~erg/s/cm$^2$. 
Compared with the $Faint~Gold$ sample, the $Bright ~Gold$ sample tend to have a much larger significant radius than the core radius, a larger number of photon within the significant radius, and a much flatter profile.
Especially, in the $\beta<2/3$ regime, the $Bright~Gold$ sample has a much smaller $\beta$ value (i.e., much flatter profile) than the $Faint~Gold$. Thus, our detection efficiency of flat clusters is higher for the bright ones, which comes from the limitation of the telescope sensitivity. In another word, our detection of such flat clusters with shallow RASS data shows that it is promising to use this algorithm to make detection of large number of very extended sources with deeper observation or the detector with better sensitivity.

\section{Discussion}
\label{sec:discussion}

\subsection{Very extended candidates}
\label{subsec:extended_candi}

\begin{figure}[t]
    \centering
    \includegraphics[width=0.4\textwidth]{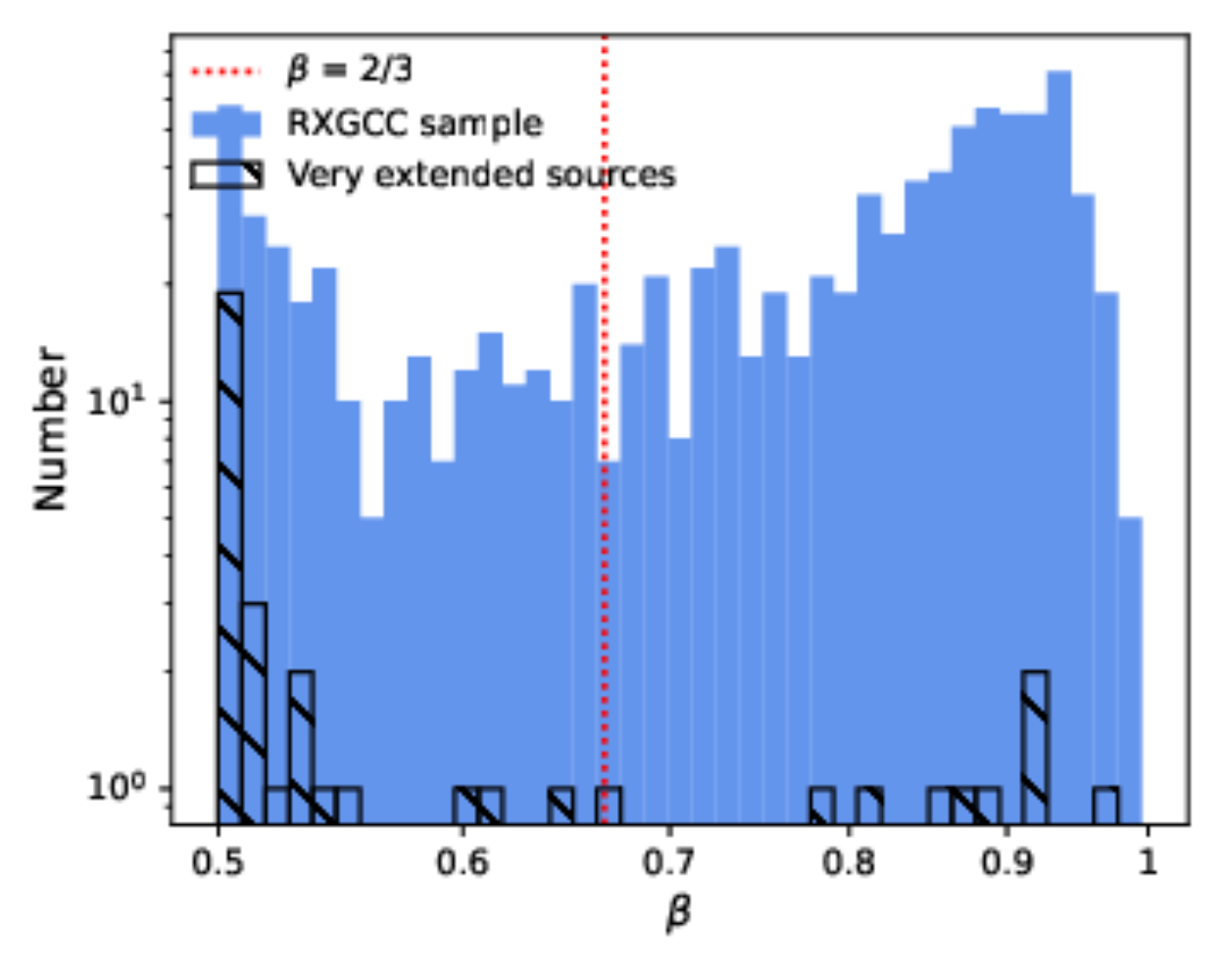} 
    \caption{The $\beta$ distribution of the whole RXGCC and the very extended sub-sample. The very extended sub-sample comprise of clusters located at the bottom-right part of the dividing line in the top-left panel of Fig.~\ref{fig:r_cr}.}
    \label{fig:beta_ext}
\end{figure}

In the top-left panel of Fig.~\ref{fig:r_cr}, we show the relation between the significant radius ($R_{\rm sig}$) and $R_{500}$. We find $R_{\rm sig}$ tends to be
larger than $R_{500}$. Thus, we separate out very extended clusters located at the
bottom-right corner, using the dotted dividing line therein. The comparison of the
$\beta$ distribution of the whole RXGCC and this very extended sub-sample is shown in Fig.~\ref{fig:beta_ext}. Most of these extended candidates have $\beta<2/3$, with a much flatter profile than the typical clusters. The detection of cluster candidates
with such flat profiles demonstrates the efficiency of our algorithm and its possibility to detect much more extended clusters with forthcoming observations.

\subsection{False detections}
\label{subsec:false_detections}

In our detections, except of $944$ cluster candidates compiled into the RXGCC catalog, 
there are $364$ false detections.
There are several possible reasons for false detections, 
such as large 
variation of exposure time, background, or the column density of 
neutral hydrogen, contamination from foreground or background, 
clustering of X-ray point sources (stars or AGNs), 
projection overlapping of outskirts of nearby bright/extended 
X-ray sources, or projection overlapping of high-redshift clusters. 

\begin{figure}[t]
\centering
\includegraphics[width=0.45\textwidth]{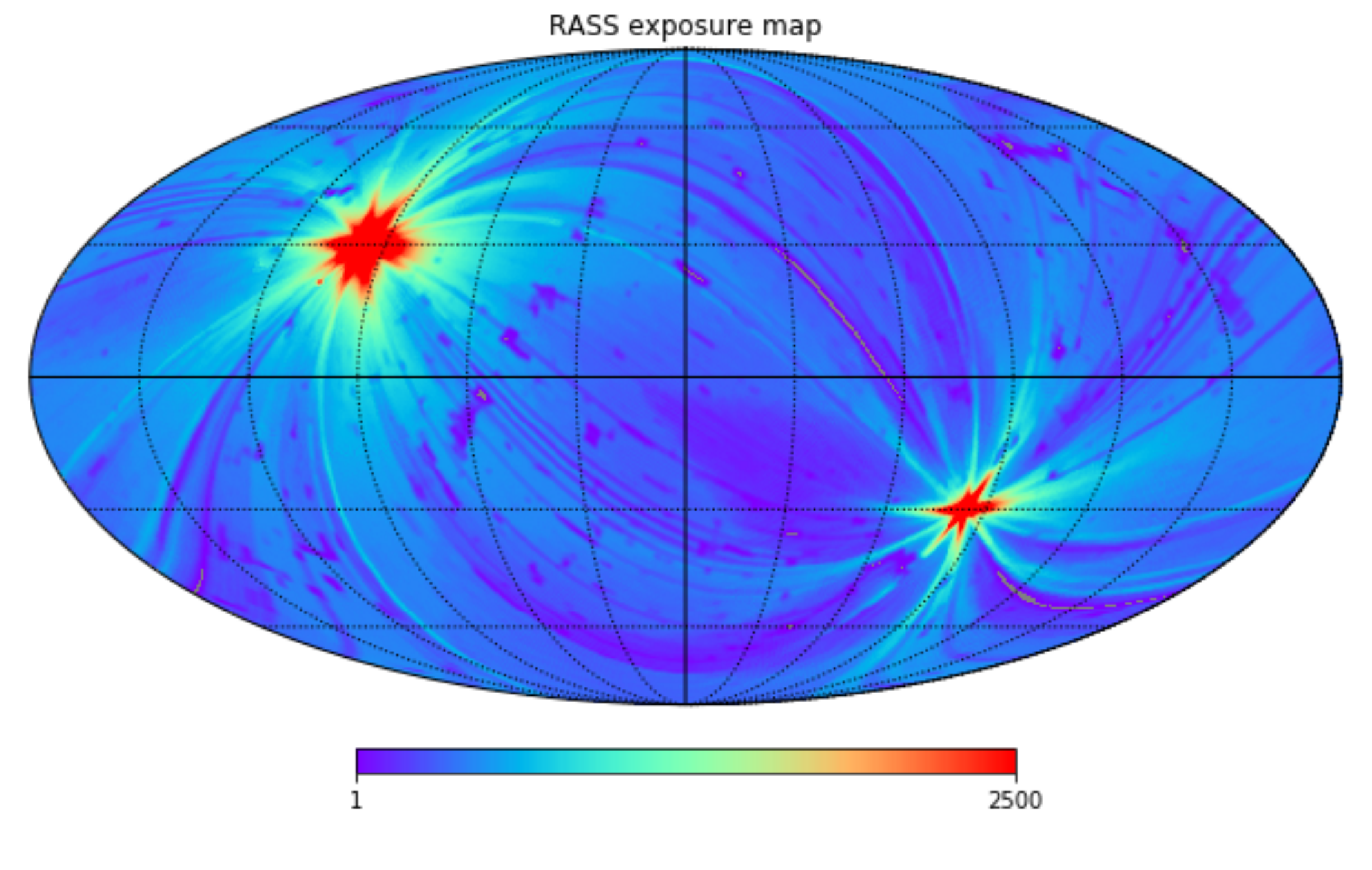}  \\
\includegraphics[width=0.45\textwidth]{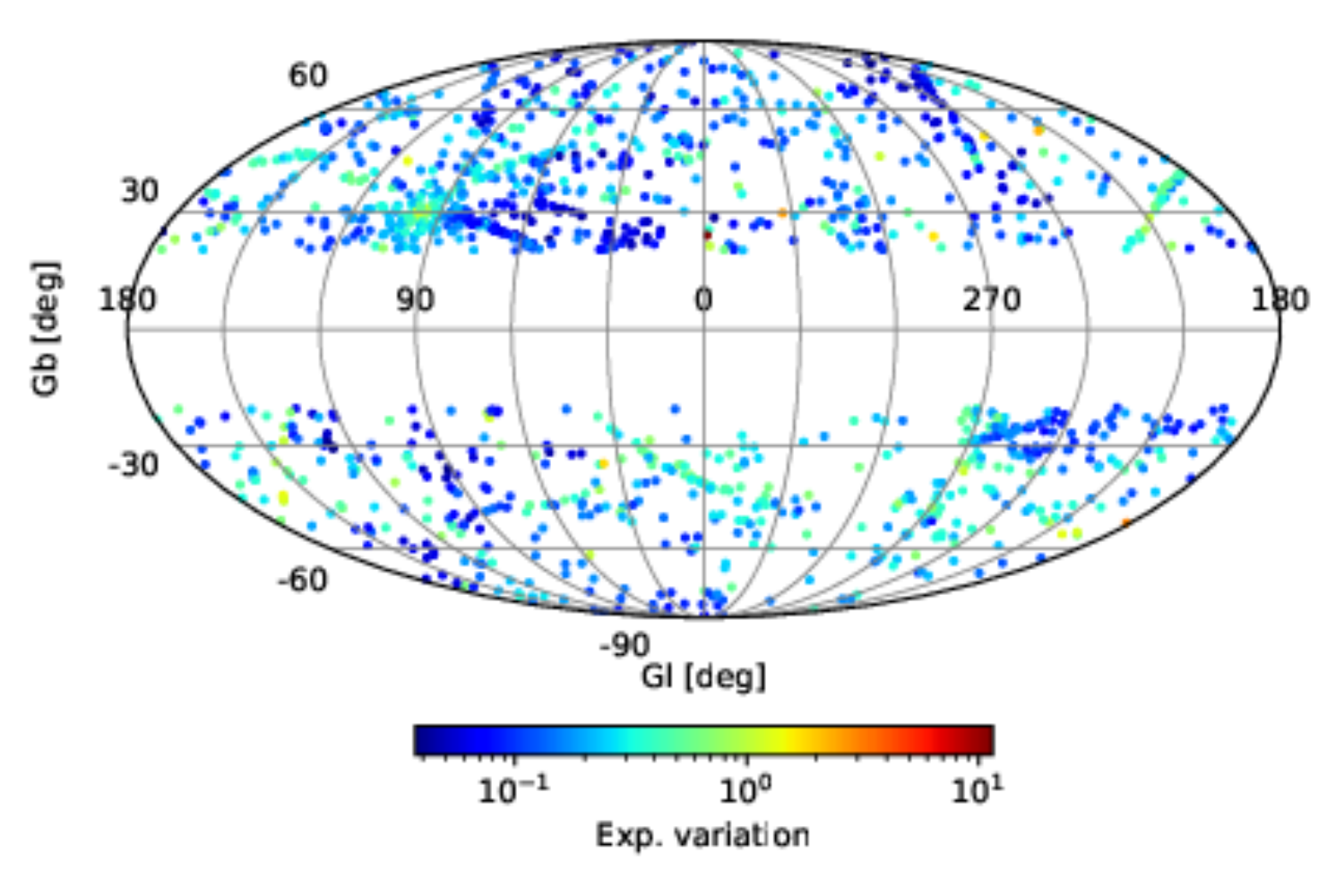} \\ 
\includegraphics[width=0.42\textwidth]{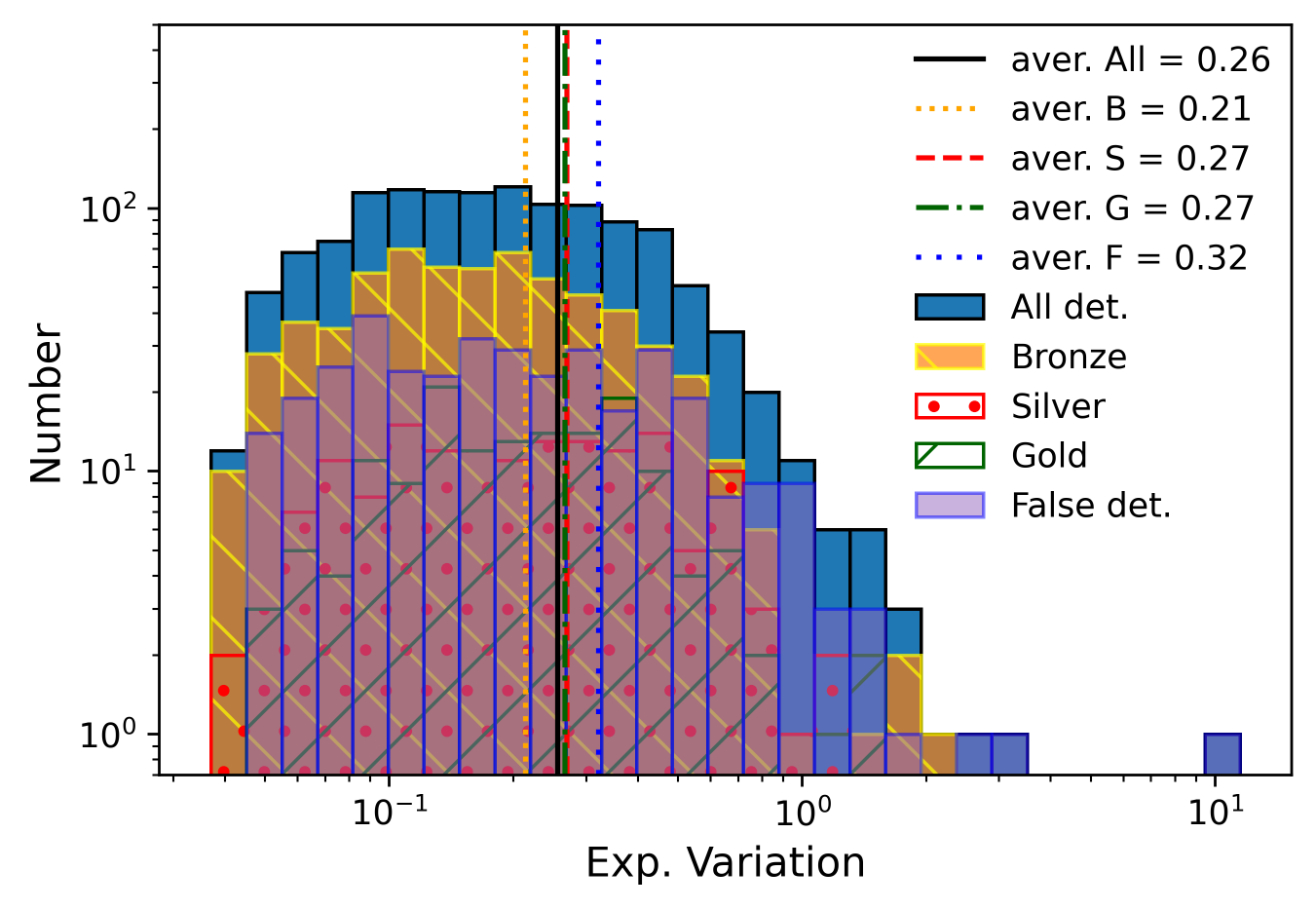}\\ 
\includegraphics[width=0.42\textwidth]{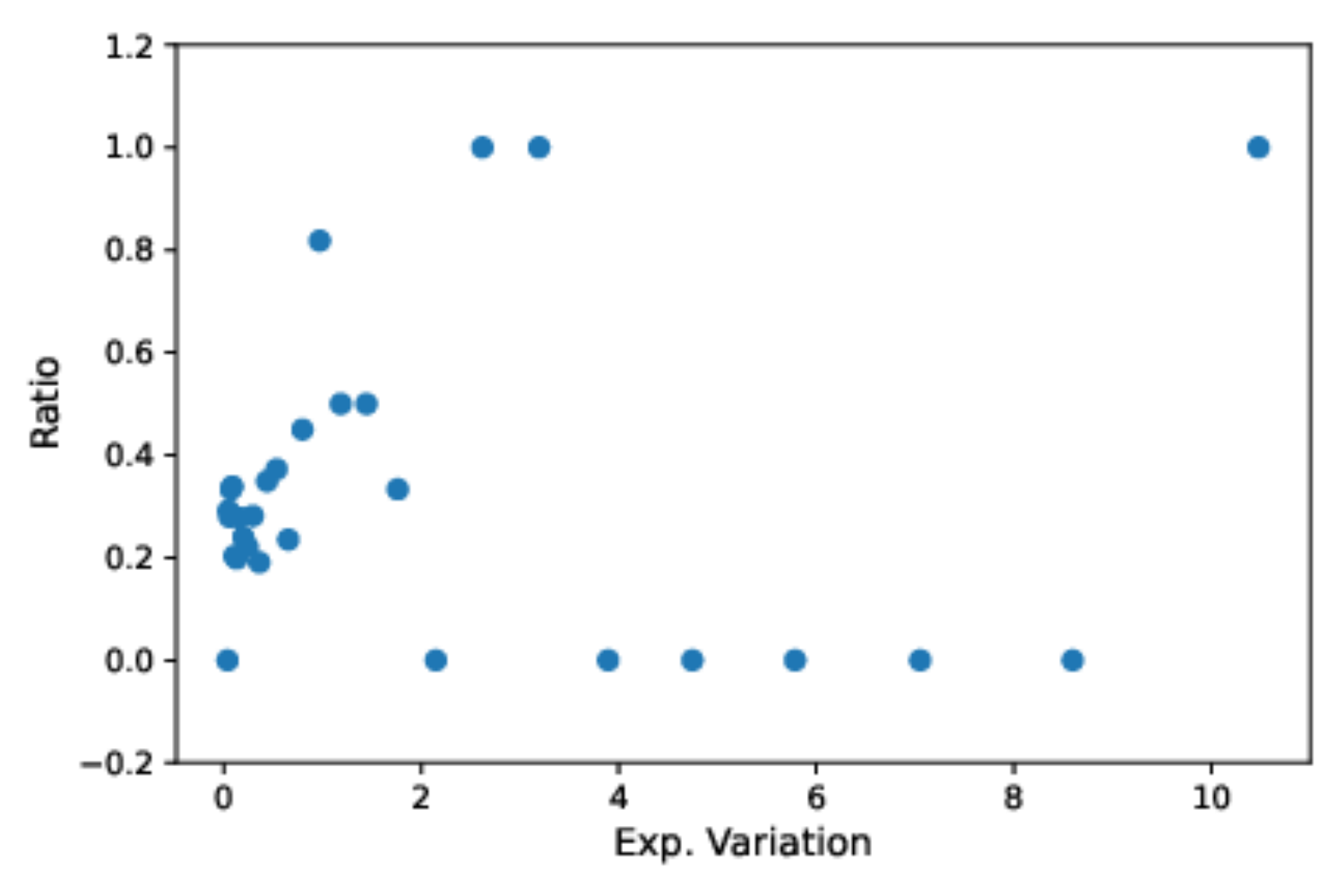} 
\caption{\footnotesize
{Top panel: the RASS-II exposure map (adapted from the Fig.~$1$ of \citealt{Voges1999}.)
Second panel: all-sky distribution of all detections, colors vary with the exposure variation. 
Third panel: the exposure variation distribution for detections in classes, the average value are overlaid with vertical lines. 
Bottom panel: the relation between the false detection ratio with exposure variation.}}
\label{fig:delta_em}
\end{figure}

The spatial distribution of false detections is 
shown with red dots in the bottom panel of Fig.~\ref{fig:allsky}. 
The ratio of false detections in our detection is $27.8\%$. 
Most of them are located at the ecliptic poles and along ecliptic 
longitudes, where the exposure variation is large (shown at the
second panel of Fig.~\ref{fig:delta_em}). The exposure variation is taken from the difference between the highest and 
lowest exposure time of pixels in $1^\circ\times 1^\circ$ region 
and divided by the median value. 
The way of RASS scanning causes large exposure variation at the direction vertical to the scanning direction, shown in the top panel of Fig.~\ref{fig:delta_em}, which is adopted from the Fig.~$1$ of \citet{Voges1999}. The spatial correlation in the top and the second panel indicates that large exposure variation is one reason for the false detections. 

In the third panel of 
Fig.~\ref{fig:delta_em}, we plot the histogram of exposure 
variation for detections in classes. It is clear the 
average exposure variation of $False~detections$ is higher than other classes and the whole detection sample.
In the bottom panel of Fig.~\ref{fig:delta_em}, it shows the positive correlation between the false detection ratio with the exposure variation. Therefore, we take the large exposure variation as the main reason for false detections. 

\subsection{Cross-match RXGCC with representative ICM-detected cluster catalogs}
\label{subsec:det_reflex_noras}

In this section, we discuss the robustness of cross-matching criteria in the catalog cross-matching, as described in Sec.~\ref{subsec:class_criteria}. These cross-matching criteria,
offset $<15'~\&$ offset $<0.5~$Mpc~$\&~\Delta z<0.01$, are used for all cross-matching in this work.

\begin{figure}[t]
\centering
\includegraphics[width=0.4\textwidth]{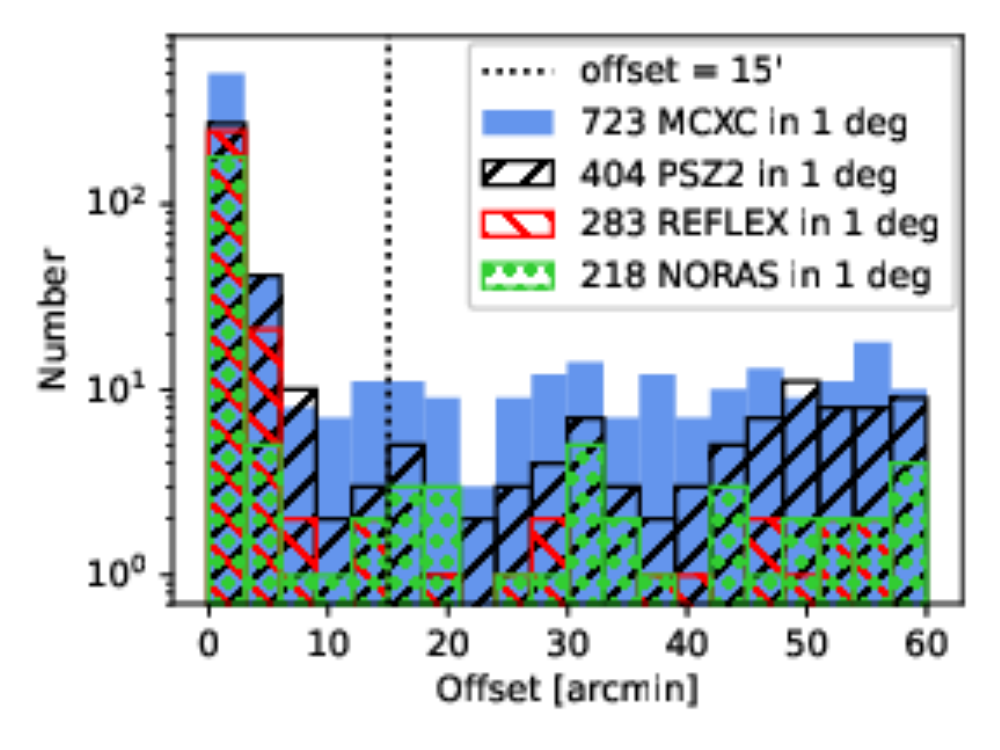} 
\includegraphics[width=0.4\textwidth]{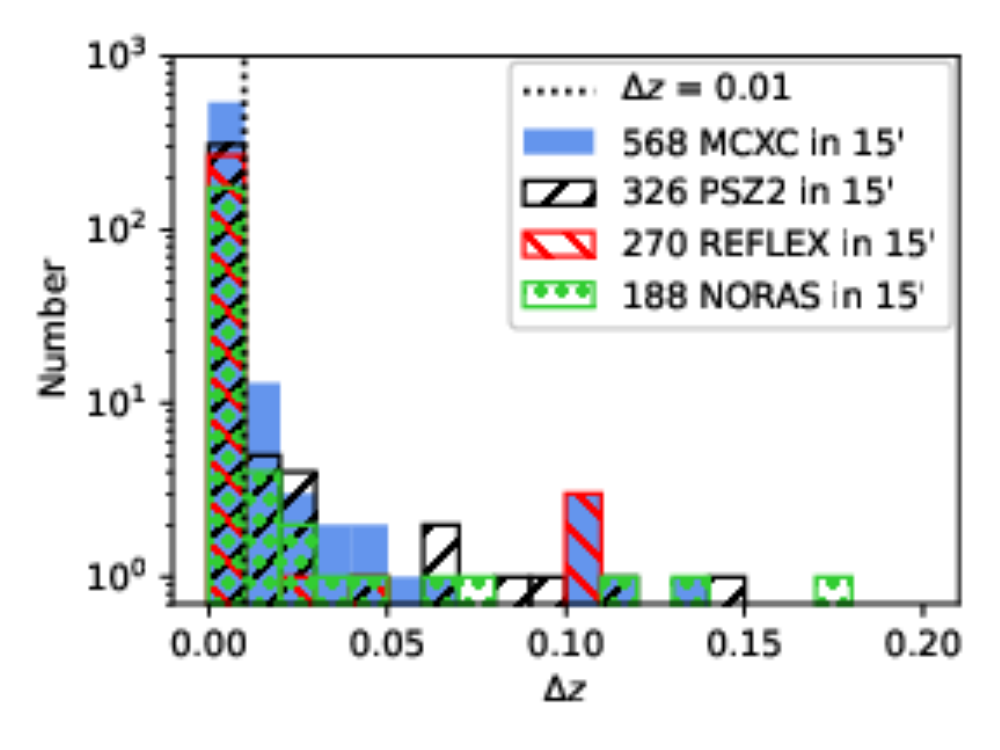} 
\caption{\footnotesize
{The difference of positions and redshifts between
our candidates and position-matched MCXC/PSZ2/REFLEX/NORAS clusters.
In the top panel, the offset threshold is set as $1$~degree,
without setting redshift limit. In the bottom panel,
the position offset is set as $<15'$. The dotted vertical lines
labels out the criteria used in this work.}}
\label{fig:dis_mp}
\end{figure}

Thus, we cross-match the RXGCC catalog with four representative ICM-detected cluster catalogs. We consider two ROSAT-based X-ray cluster catalogs, the ROSAT-ESO flux limited X-ray catalog (REFLEX, \citealt{Bohringer2004}) and the northern ROSAT all-sky catalog (NORAS, \citealt{Bohringer2000}), as well as the MCXC and PSZ2 catalogs to represent large cluster 
catalogs in X-ray and microwave bands. Firstly, we consider all matches with offset $<1$~degree, and show the offset distribution. Secondly, we constrain the offset within $15'$, and show the redshift difference of matches. The result is shown in Fig.~\ref{fig:dis_mp}. When the offset increases, the number of matching pairs decreases firstly, and increases later for the random distribution and projection effect.
Similarly, with the redshift difference increases, the number of matching pairs has an exponential decrease, and a tiny increase later.
The criteria labeled with dotted vertical lines are used in our work to ensure a high probability for a  physical correlation of matched clusters and a low probability of projection effects.

\subsection{Known X-ray sources}
\label{X ray sources}

\begin{figure}[t]
\centering
\includegraphics[width=0.38\textwidth]{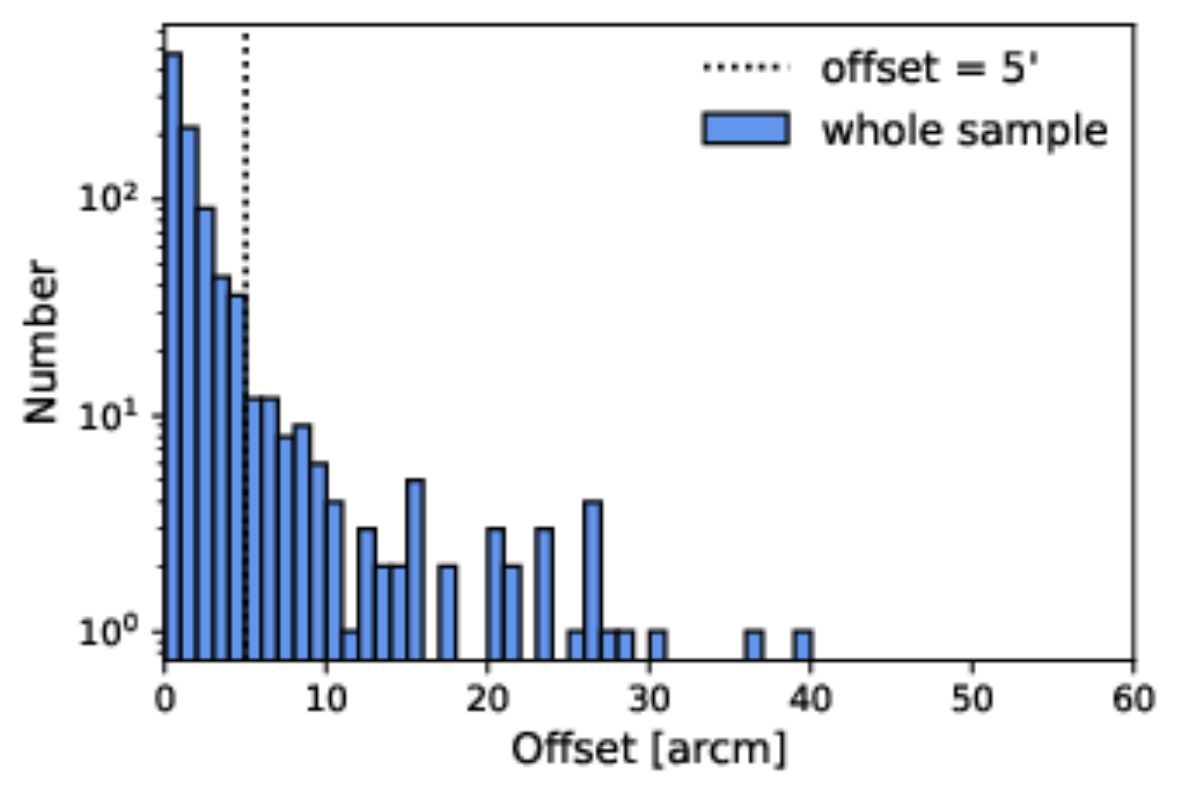} 
\caption{\footnotesize
{The offset distribution between RXGCC clusters and the nearest RXS source.}}
\label{fig:dis_rxs}
\end{figure}

In this section, we cross-match the RXGCC catalog with the ROSAT
X-ray Source Catalog (RXS, \citealt{Voges1999,Voges2000,Boller2016}).
The RXS is often used as the underlying resource of possible
X-ray clusters \citep[e.g.,][]{Bohringer2000,Bohringer2004}. So, 
in case of RXGCC clusters not included in
the RXS, they will be missed by all cluster searching projects
which take the RXS catalog as input. The ratio of RXGCC clusters
with RXS detection in $5'$ is $91.1\%$, as shown in
Fig.~\ref{fig:dis_rxs}. That is, there are $84$ clusters without 
any RXS sources within $5'$. These clusters are definitely 
missed by X-ray cluster detection projects starting from the RXS sources.

\section{Conclusion}
\label{sec:conclusion}
In this project, we set out to reanalyze the RASS in the [$0.5-2.0$]~keV band and search for possibly missed X-ray clusters.
With the wavelet filtering, source extraction, and maximum likelihood fitting, 
we make $1308$ detections. Combining 
the RASS images 
with optical, infrared, micro-wave band images, the neutral hydrogen distribution, and the spatial and redshift distribution of galaxies in the field, we identify $944$ clusters and remove other false detections. 
Among these clusters, there are $149$ ones detected for the first time, and additional $154$ ones detected through the ICM emission for the first time. 
For each candidate, we 
estimate the redshift using the distribution of the spectroscopic and photometric 
redshift of galaxies, and use the growth curve analysis to estimate the parameters, such as count 
rate, flux, luminosity and mass. 

$115$ of the $149+154$ clusters detected here with ICM emission for the first time have fluxes $>3\times10^{-12}$ erg/s/cm$^2$; i.e., above the rough flux limits applied in previous RASS cluster catalogs. We find that the new clusters deviate in their properties from the previously known cluster population. In particular, the new clusters have flatter surface brightness profiles, which makes their detection more difficult with detection algorithms optimized for point sources. This new cluster population now needs to be followed up with deep pointed observations with XMM-Newton and Chandra in order to determine their properties in more detail. After that, it allows us to estimate the magnitude of any potential bias in the estimated completeness of previous surveys, as well as its impact on $\Omega_{\rm M}$ and $\sigma_8$ constraints. Furthermore, these findings will help the upcoming cluster searches with eROSITA data.

\begin{acknowledgements}
We acknowledge support from the National Key R$\&$D Program of China (2016YFA0400703) and the National Science Foundation of China (11721303, 11890693).
The authors wish to thank Angus Wright, Jens Erler, 
Cosmos C. Yeh, Chaoli Zhang, Aiyuan Yang, Linhua Jiang, Zhonglue Wen,
and Jinlin Han for useful help and discussions during the 
development of this paper. The authors 
thank the support from the CAS-DAAD Joint Fellowship 
Programme for Doctoral Students of Chinese Academy of 
Sciences (ST 34). WX acknowledges the support of the 
Chinese Academy of Sciences through grant No. XDB23040100 
from the Strategic Priority Research Program and that of 
the National Natural Science Foundation of China with grant No. 11333005. 
We acknowledge support from the Chinese Academy 
of Sciences (CAS) through a China-Chile Joint Research Fund 
\#1503 administered by the CAS South America Center for Astronomy.
\end{acknowledgements}

\bibliographystyle{aa} 
\bibliography{main.bbl} 

\appendix

\section{Method comparison with Paper I}
\label{sec:diff_paper1}

Compared with Paper I, there are some changes in the method.
\begin{itemize}
\item 1. The $40$ cluster candidates in \citet{Takey2016} are discarded since they do not provide spectroscopic redshift estimations. The \citet{Wen2012} was updated in \citet{Wen2015}, 
and the updated catalog is used in this work. 
The X-ray clusters in \citet{Wen2018} are not considered
as ICM-detected clusters, because they are detected with 
optical and infrared data. In addition, more recent
published X-ray cluster catalogs are taken into account.
\item 2. The galaxy redshift is constrained to the 
range $0<z<0.4$. Since the depth of RASS makes
it difficult to detect the X-ray emission from extended clusters
at higher redshift. The inclusion of high redshift galaxies will 
bring unnecessary contamination for the redshift determination. 
\item 3. The cross-matching with previously identified clusters 
is performed with the offset both $<0.5$~Mpc and $<15'$, 
instead of $<15'$ in Paper I. This comes from the physical 
size corresponding to $15'$
varies largely with the redshift, which can be largely different from the 
typical cluster size. Thus, the previous offset threshold casts doubt on 
the validity of the corresponding cross-matching, and further classification.
\item 4. The spectroscopic and photometric redshifts of galaxies 
are combined, instead of taken separately in Paper I, 
to make the redshift estimation. 
This comes from the fact that the combination of galaxy redshift
information provide a more accurate estimation, especially
in the area with only a limited number of spectroscopic galaxy redshift
are available.
\end{itemize}

\section{Estimation of contamination ratio}
\label{sec:contam}

Optical and infrared cluster catalogues contain larger number of 
objects in comparison with ICM-based cluster catalogues.
In our classification, the cross-matching
of our detections with previous optical/infrared-identified clusters might result in mis-identification due to projection effects.
Taking the catalog from \citet{Wen2012} as an example, there are $132\,684$
clusters identified within the SDSS-III area of $14\,000$ deg$^2$.
That is, there is on average $\sim1$ cluster in the circular area with the
radius of $11$~arcmin. And $11$~arcmin is $\sim 1$~Mpc for a cluster with
$z=0.0770$, which is the median redshift of the RXGCC sample.
This means, at least in some area (e.g., within the SDSS coverage),
there is on average one optical cluster within an area with the size of a
typical cluster. Although our algorithm of redshift determination and
visual check will help to remove much of
this kind of contamination, some contamination is expected to exist.

To estimate the contamination ratio from the projection effect,
we make $200$ simulations. In each simulation, there are
$1100\sim1200$ positions taken randomly with the Aitoff projection. We also
discard some areas from the whole sky, as described in Sec.~\ref{subsec:data}.
Then, we classify them by cross-matching them with previous-identified clusters, as described in Sec.~\ref{subsec:cluster-literature}. 
The classification criteria in Sec.\ref{subsec:class_criteria} are taken, while no visual check is used for these simulation detections.
In each simulation, the contamination number and ratio are
normalized to $1308$ total detections.

\begin{figure}[t]
\centering
\includegraphics[width=0.5\textwidth]{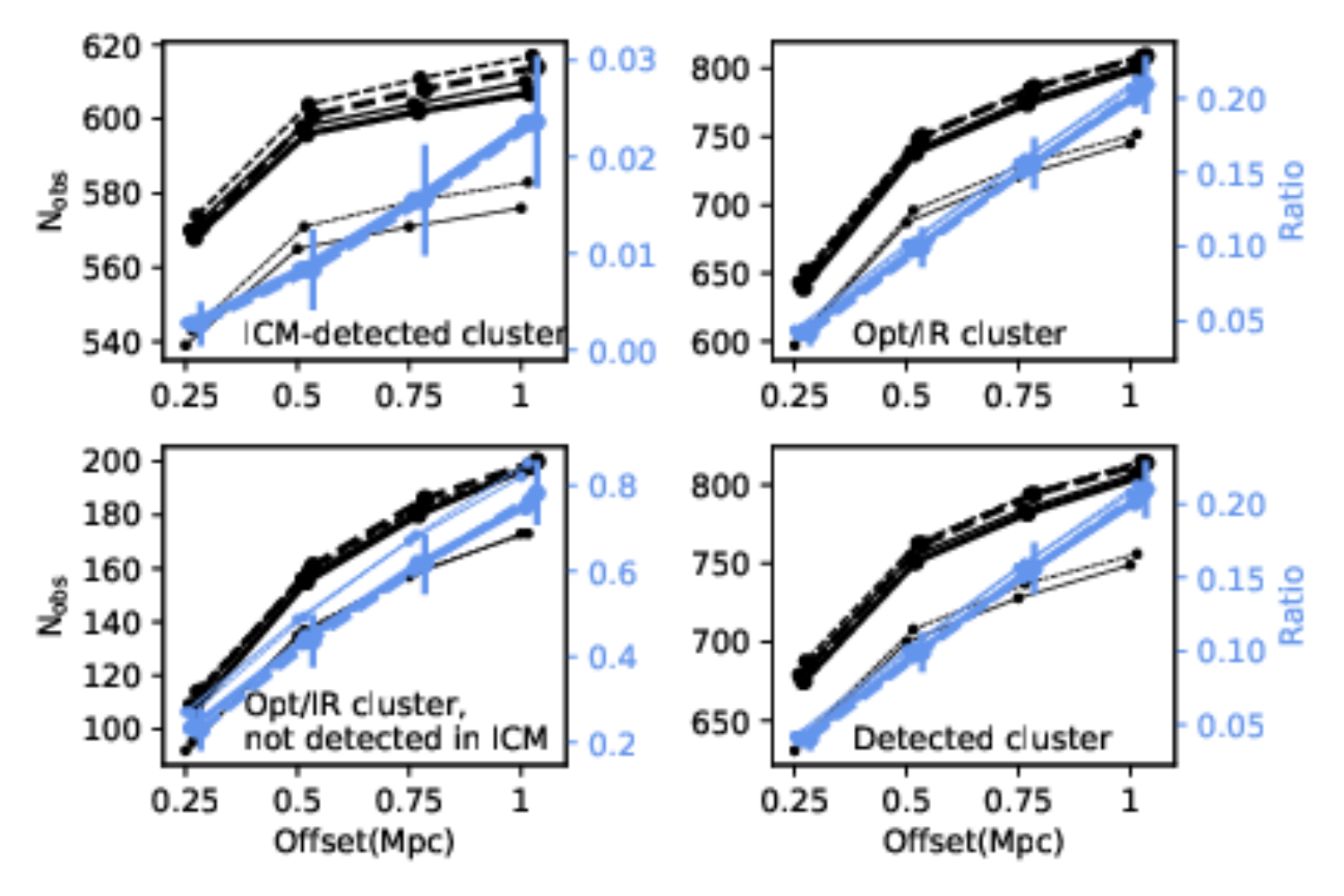} 
\caption{\footnotesize
{The number of detections (shown as black line with labels on the left) and contamination ratio (shown as blue line with labels on the right)
varies with the offset threshold in the cross-matching with previously identified clusters (shown in X-axis).
The offset has been set to be $0.25$, $0.5$, $0.75$, $1.0$~Mpc. The dots in the plot are shifted to the right with a little distance to avoid the overlapping. 
The detection numbers and contamination ratio for previous ICM-detected clusters (top-left panel), previous optical/infrared clusters (top-right panel), previous optical/infrared clusters not detected in ICM (bottom-left panel), the whole sample of detections (bottom-right panel) are shown. 
The solid and dashed lines shows the results when the upper limit of galaxy redshift is set to be $0.3$ and $0.4$, respectively. 
To estimate the redshift of detections, we consider three different situations. Firstly, we only take spectroscopic galaxy redshift into account (shown with thin lines). Secondly, we consider the spectroscopic galaxy redshift and photometric one separately (shown with lines in the normal width). Lastly, we combine the spectroscopic galaxy redshifts and photometric ones together without distinguishing them (shown with thick lines).}}
\label{fig:contamination_ratio}
\end{figure}

In Fig.~\ref{fig:contamination_ratio}, we show simulation results.
The upper limit of galaxy redshift is set to be $0.3$ and $0.4$ (shown with solid and dashed lines respectively in the figure),
to remove the contamination from the high redshift galaxies in deep surveys, 
which is unlikely to detect with RASS. Additionally, the redshift of galaxies
are taken in three different ways. Firstly, only 
spectroscopic redshift of galaxies is taken to make redshift estimation
(shown with thin lines). Secondly, both the 
spectroscopic redshift and photometric redshift of galaxies are taken separately 
(shown with lines in the normal width). Thirdly, both the 
spectroscopic redshift and photometric redshift of galaxies are taken combined (shown with thick lines). This way, we get the detection number
(shown with black lines), and the corresponding contamination ratio (shown with the blue lines) in simulations.
They vary with the offset threshold (shown as the X-axis) in classes.

From the figure, 
the best criteria are obtained for a low contamination ratio and a high 
detection number. Thus, both the spectroscopic and photometric redshifts 
of galaxies are combined to derive the candidate redshift 
when $0<z<0.4$, and $0.5$~Mpc $\&$ $15$ arcmin is set as offset threshold when cross-matching 
the detections with previous-identified clusters.
This way, the contamination ratio 
for ICM-detected clusters and optical/infrared-identified clusters obtained, as $0.008$ and $0.100$, respectively.

\section{Comparison of detected and undetected MCXC, PSZ2, Abell clusters}
\label{subsec:det_undet_mcplab}

\begin{figure*}[t]
\centering
\includegraphics[width=0.4\textwidth]{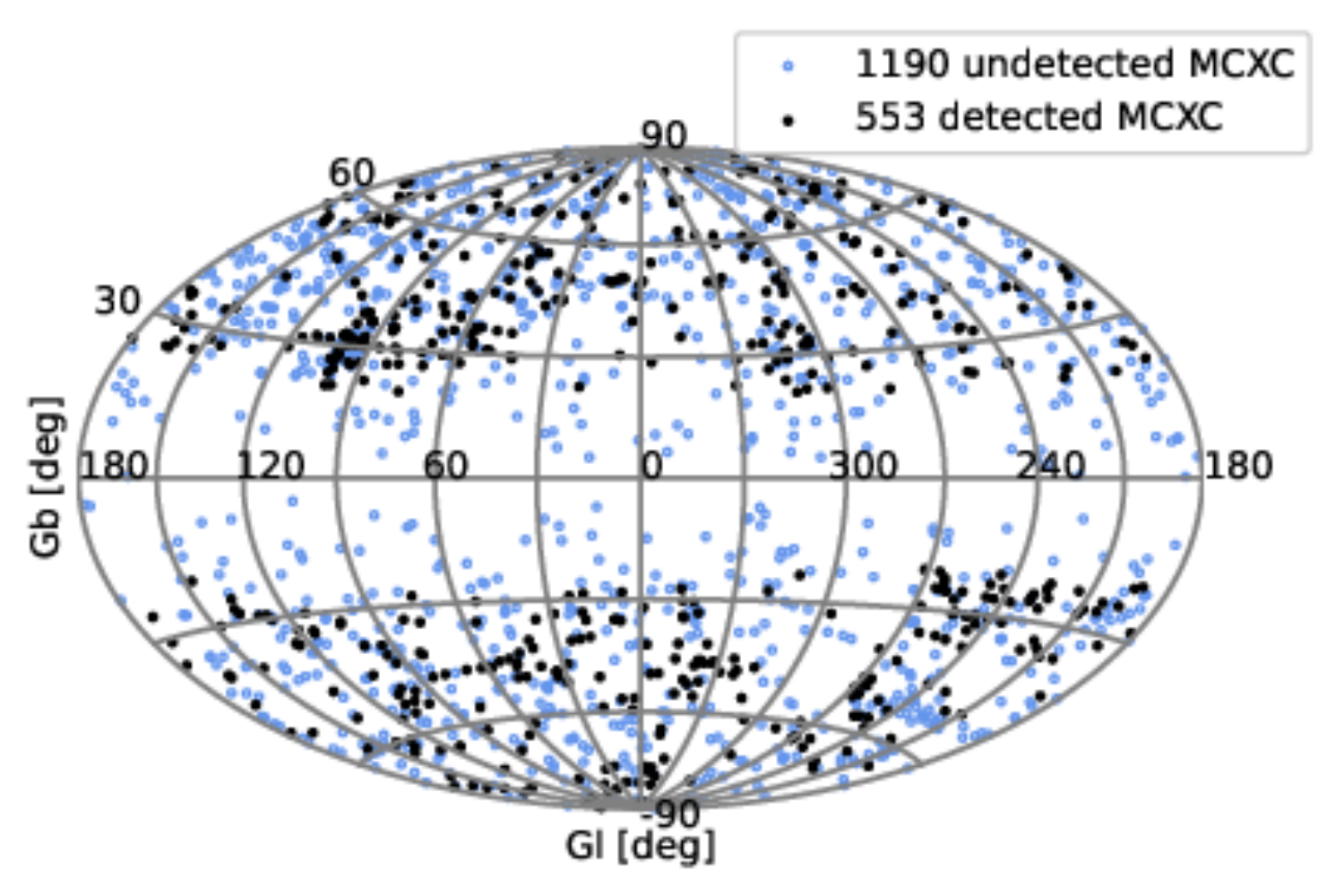} 
\includegraphics[width=0.4\textwidth]{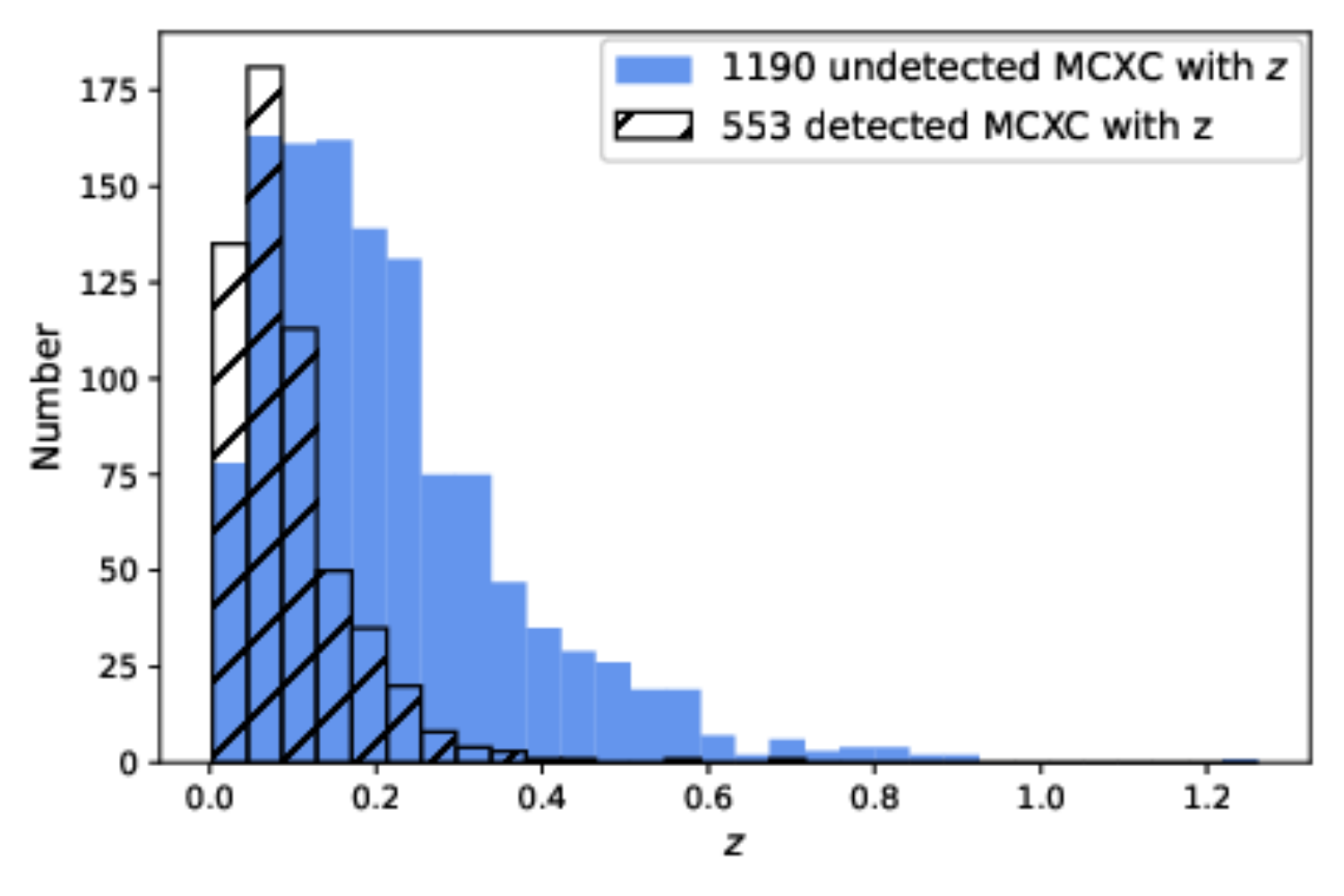} \\ 
\includegraphics[width=0.4\textwidth]{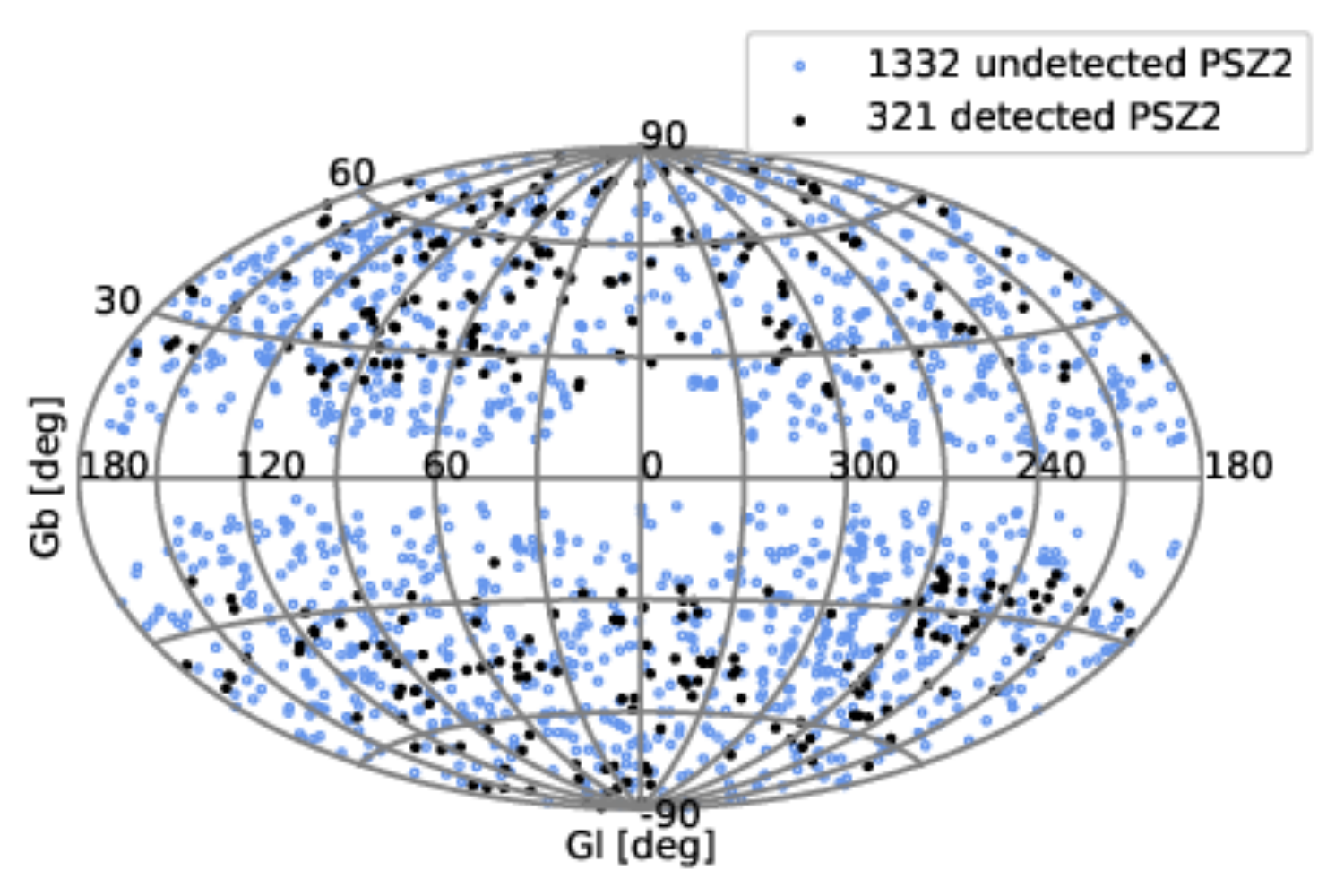} 
\includegraphics[width=0.4\textwidth]{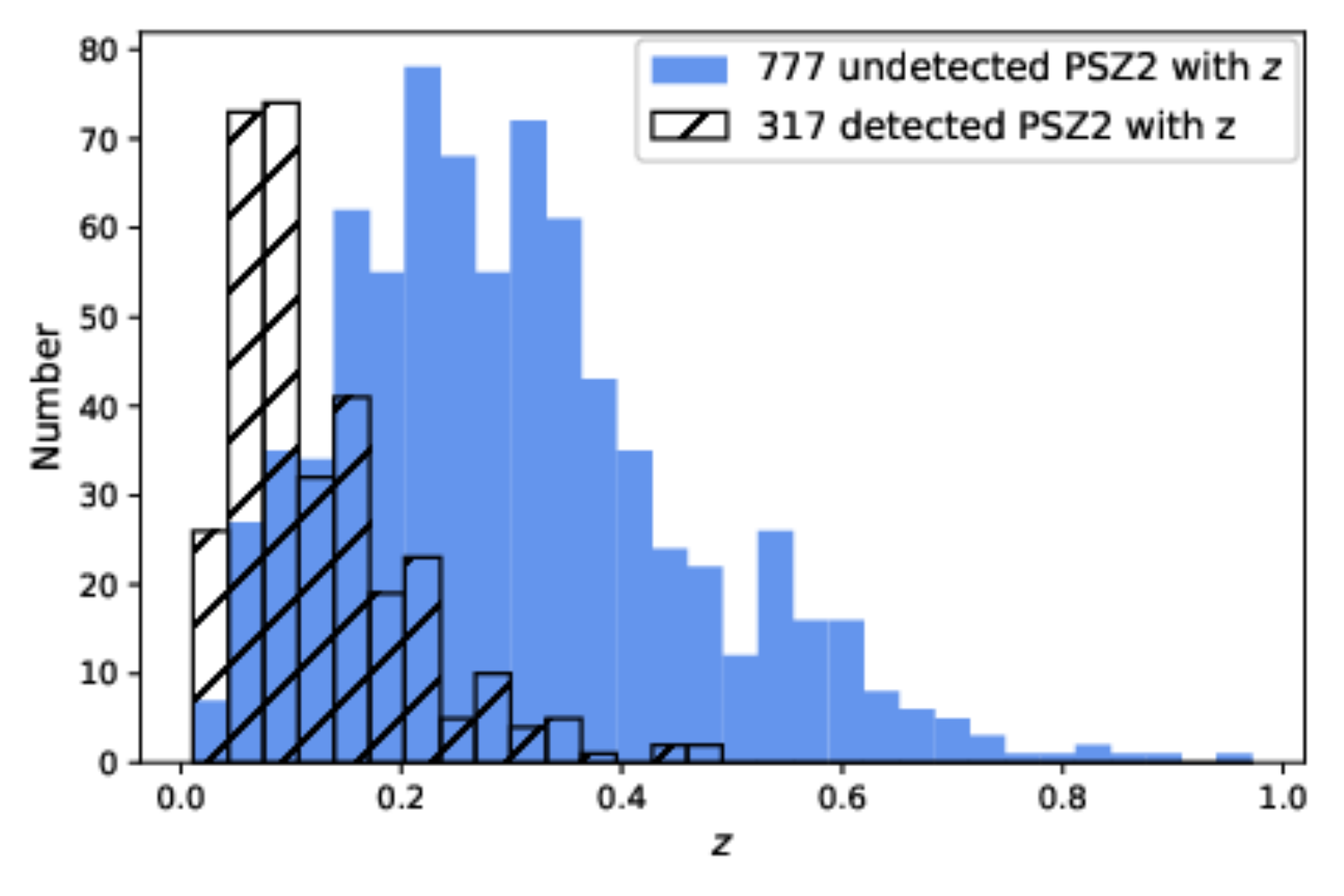}\\  
\includegraphics[width=0.4\textwidth]{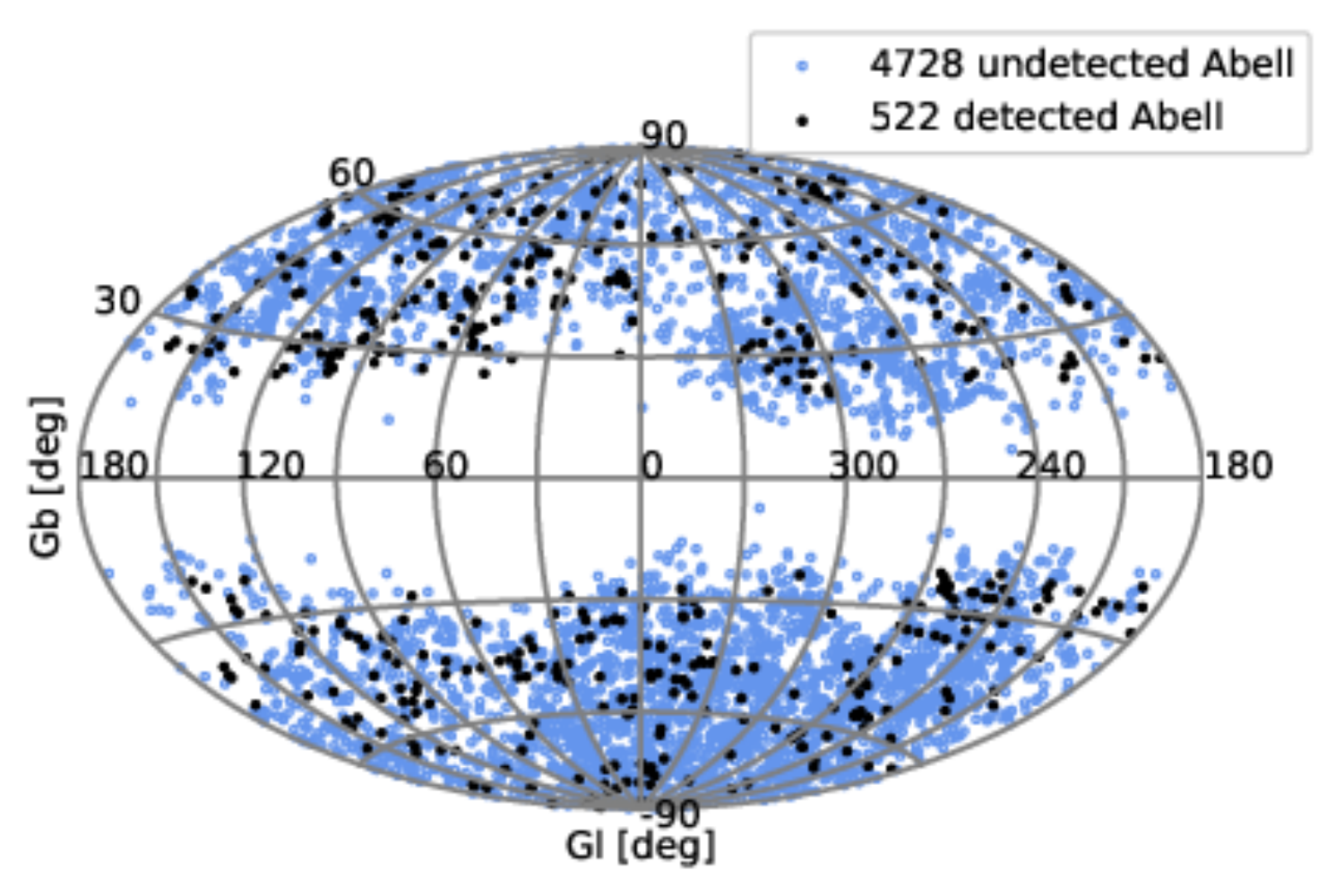} 
\includegraphics[width=0.4\textwidth]{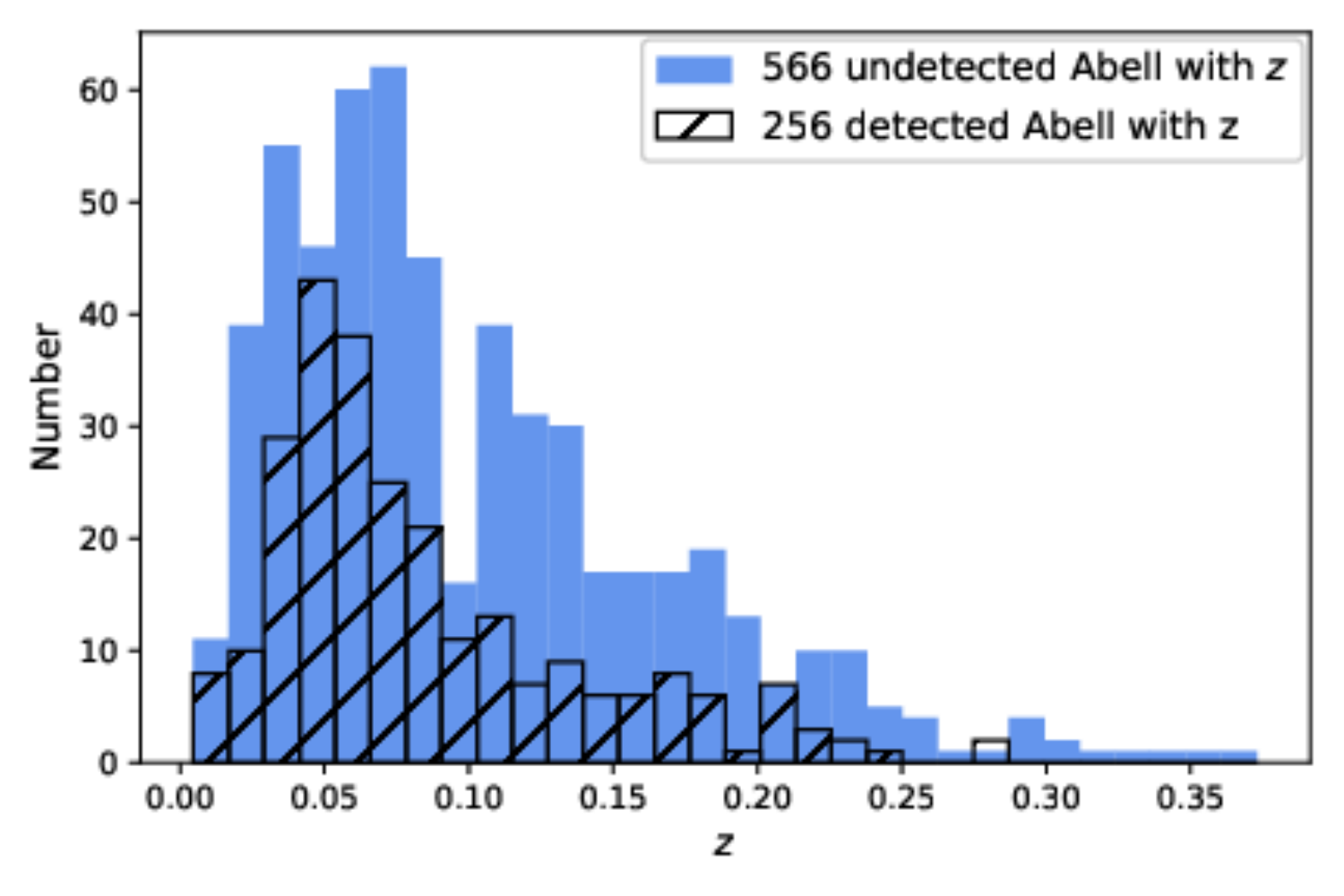} 
\caption{\footnotesize
{Comparison of spatial distribution (left panels) and redshift distribution (right panels) of detected 
and undetected MCXC, PSZ2 and Abell clusters, from the top to the bottom panels.}}
\label{fig:allsky_mc}
\end{figure*}

In this section, we compare the spatial and redshift distribution of 
detected and undetected MCXC, PSZ2, Abell clusters, as shown in Fig.~\ref{fig:allsky_mc}.
In left panels, we show the detected and undetected MCXC, PSZ2, and Abell clusters in black dots and 
blue empty circles, respectively. We hold that our algorithm has no spatial 
preference, except for the high detection efficiency in the 
ecliptic poles.
In right panels, the redshift distribution of detected and undetected MCXC, PSZ2, and Abell clusters are shown. It is obvious the detection 
efficiency peaks at $z\sim0.1$ and decreases largely 
with $z>0.2$, except for the result of Abell clusters, which is limited by the catalog itself. 

Therefore, our missing of these clusters likely comes 
from the restraints of the RASS observation. The data 
for MCXC includes not only RASS observations, but also 
ROSAT pointings with larger exposure times, which is 
vital for the detection of fainter clusters at high 
redshift. In addition, the SZ effect in microwave band is a
more efficient indicator for clusters at high redshift. 
However, the aim of this work is 
to detect missed very extended clusters, instead of a complete cluster catalog.

\section{Parameter comparison of detected and undetected REFLEX and NORAS clusters}
\label{sec:reflex_noras}

\begin{figure}[t]
    \centering
    \includegraphics[width=0.5\textwidth]{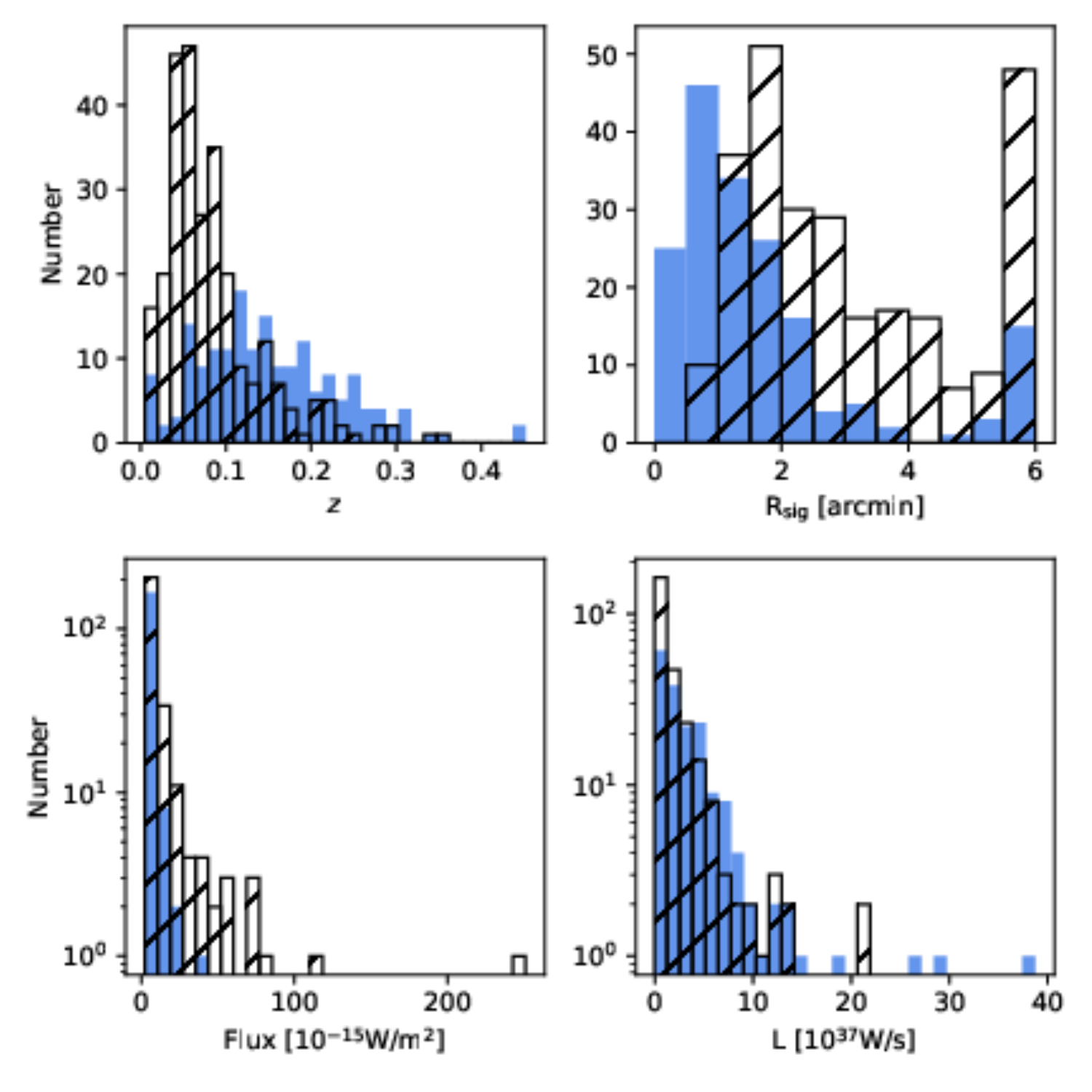} 
    \caption{The distribution of the redshift, significant radius, flux, luminosity for detected and undetected REFLEX clusters. The histogram filled with slash lines is for the detected clusters, while the blue filled histogram for the undetected ones.}
    \label{fig:det_reflex}
\end{figure}

\begin{figure}[t]
    \centering
    \includegraphics[width=0.5\textwidth]{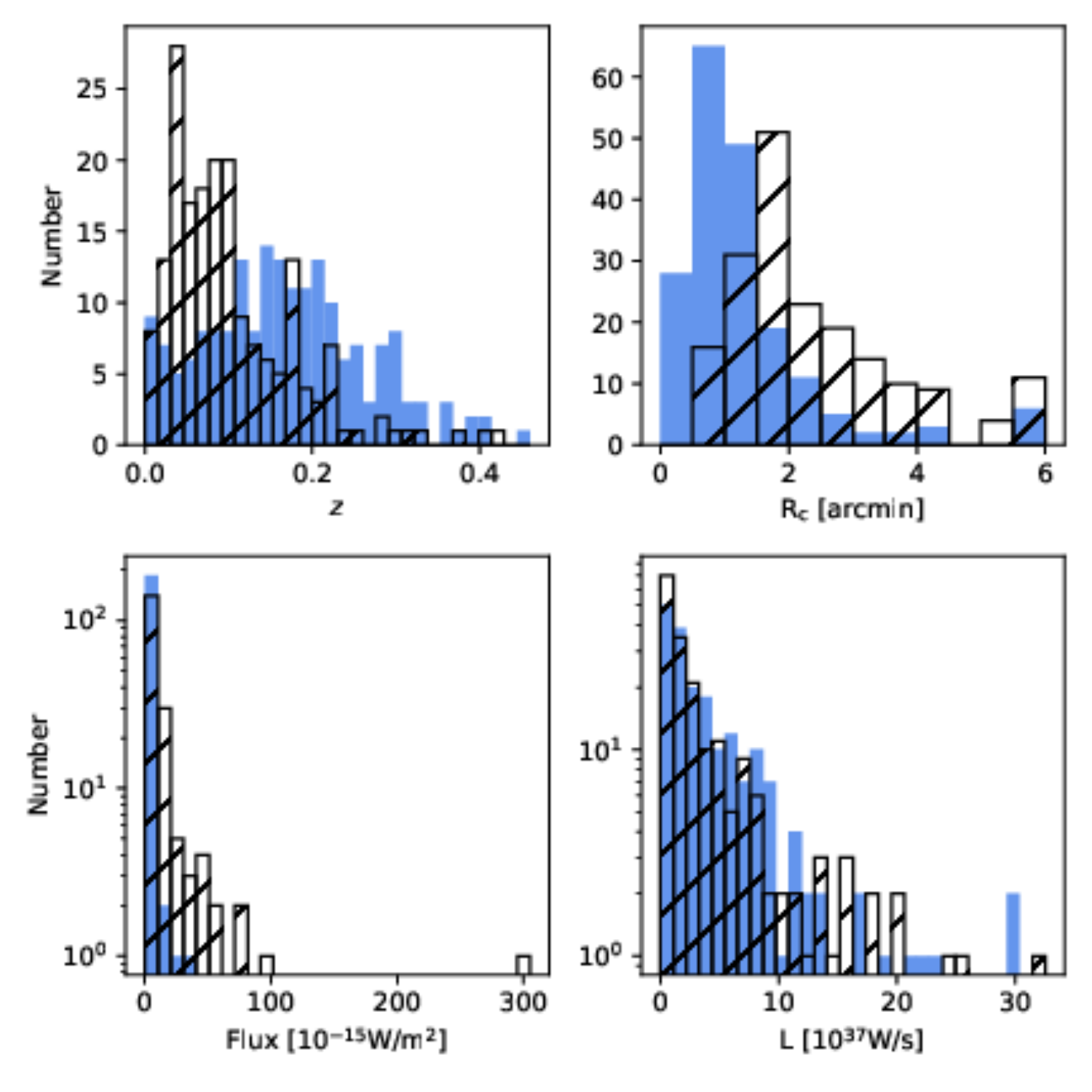} 
    \caption{The distribution of the redshift, core radius, flux, luminosity for detected and undetected NORAS clusters. The histogram filled with slash lines is for the detected clusters, while the blue filled histogram for the undetected ones.}
    \label{fig:det_noras}
\end{figure}

In Fig.~\ref{fig:det_reflex} and Fig.~\ref{fig:det_noras}, we show the distribution of the redshift, radius, flux, and luminosity for detected and undetected REFLEX and NORAS clusters. Compared with undetected REFLEX/NORAS clusters, the detected clusters tend to have lower redshift, large size, brighter flux, and lower X-ray luminosity (i.e. lower mass).

\section{Examples}
\label{sec:examples}

There are $60$ clusters with $extent>15'$ in our catalog.
For some of the detections, the $extent$ might be over-estimated. 
This mainly comes from the effects of the foreground or 
background, or the fluctuation of the exposure map. However, 
this effect can be corrected with the growth curve 
analysis (Sec.~\ref{subsec:para_estimation}). And the $R_{500}$ is more accurate
to characterize the cluster.

\begin{figure}[t]
\centering
\tiny
\includegraphics[width=0.19\textwidth]{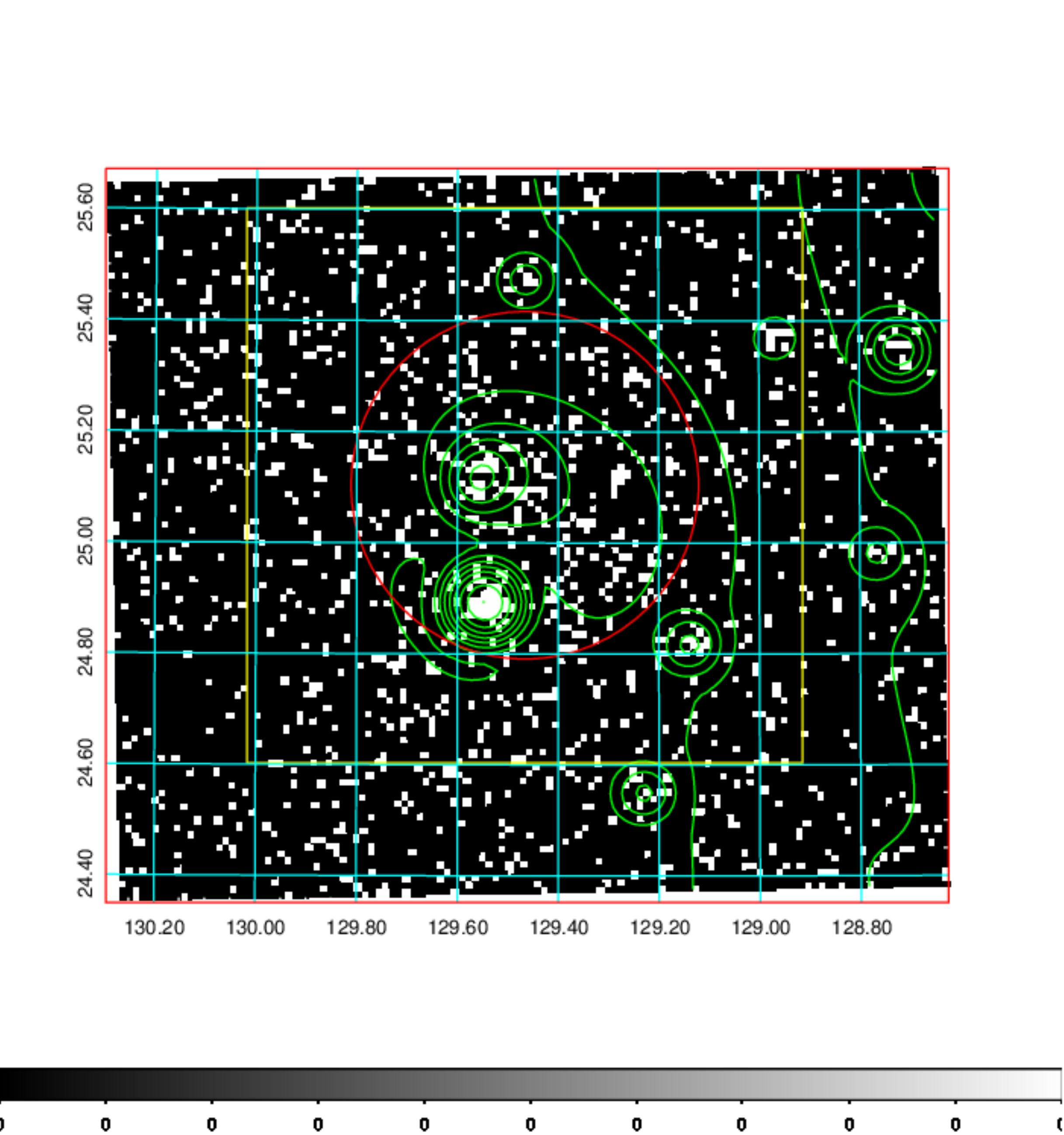} 
\includegraphics[width=0.19\textwidth]{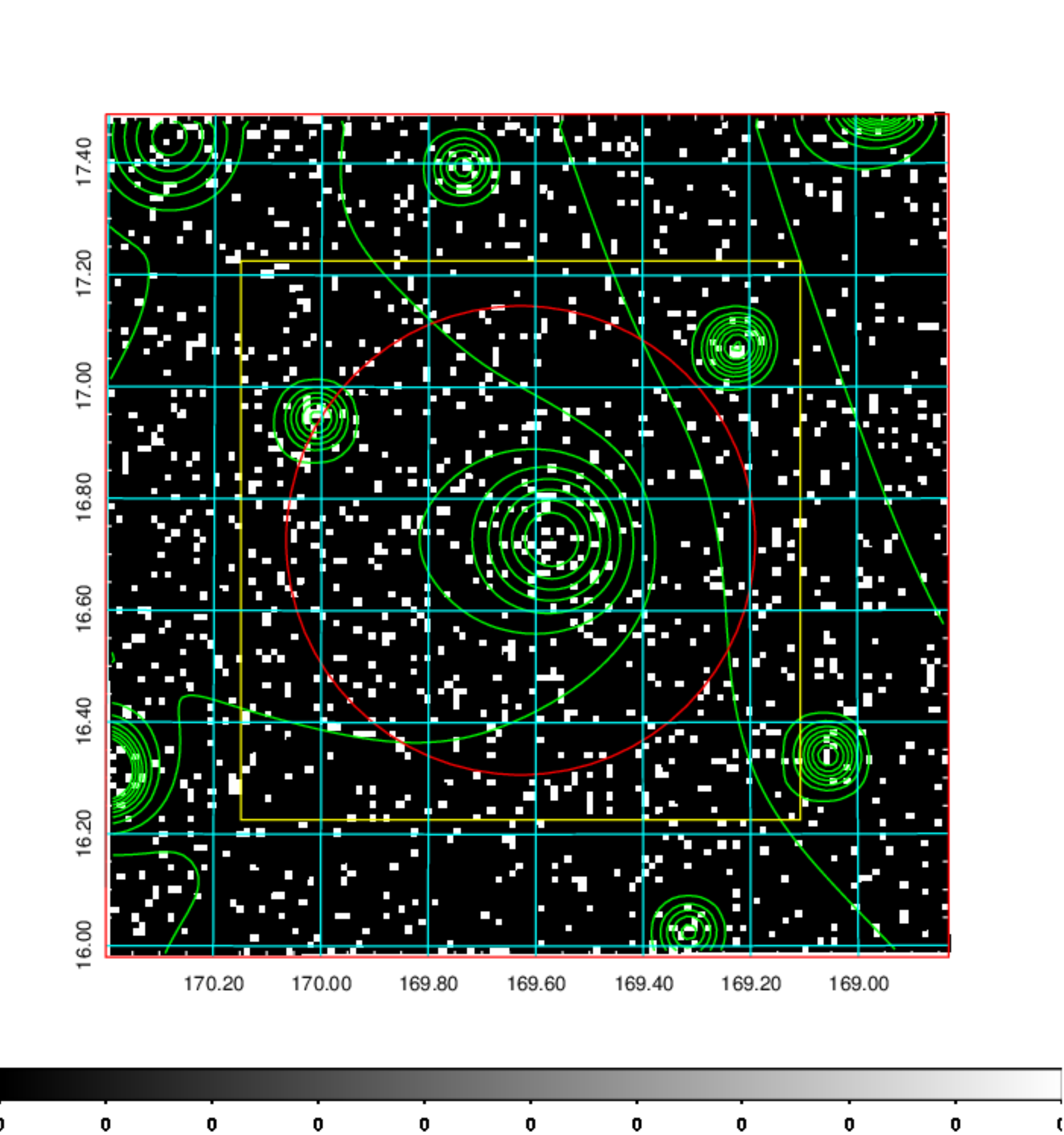} \\
\includegraphics[width=0.19\textwidth]{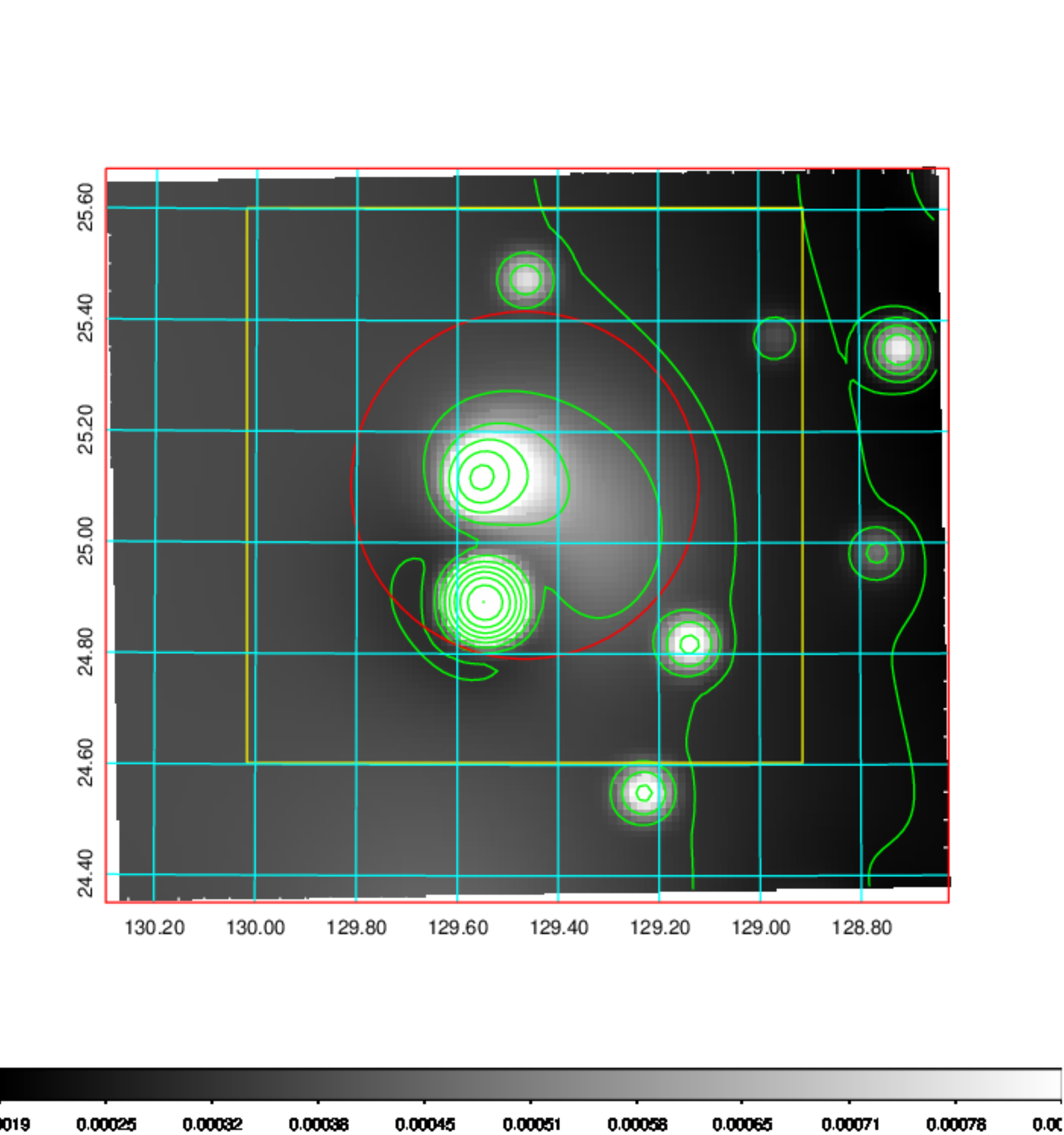} 
\includegraphics[width=0.19\textwidth]{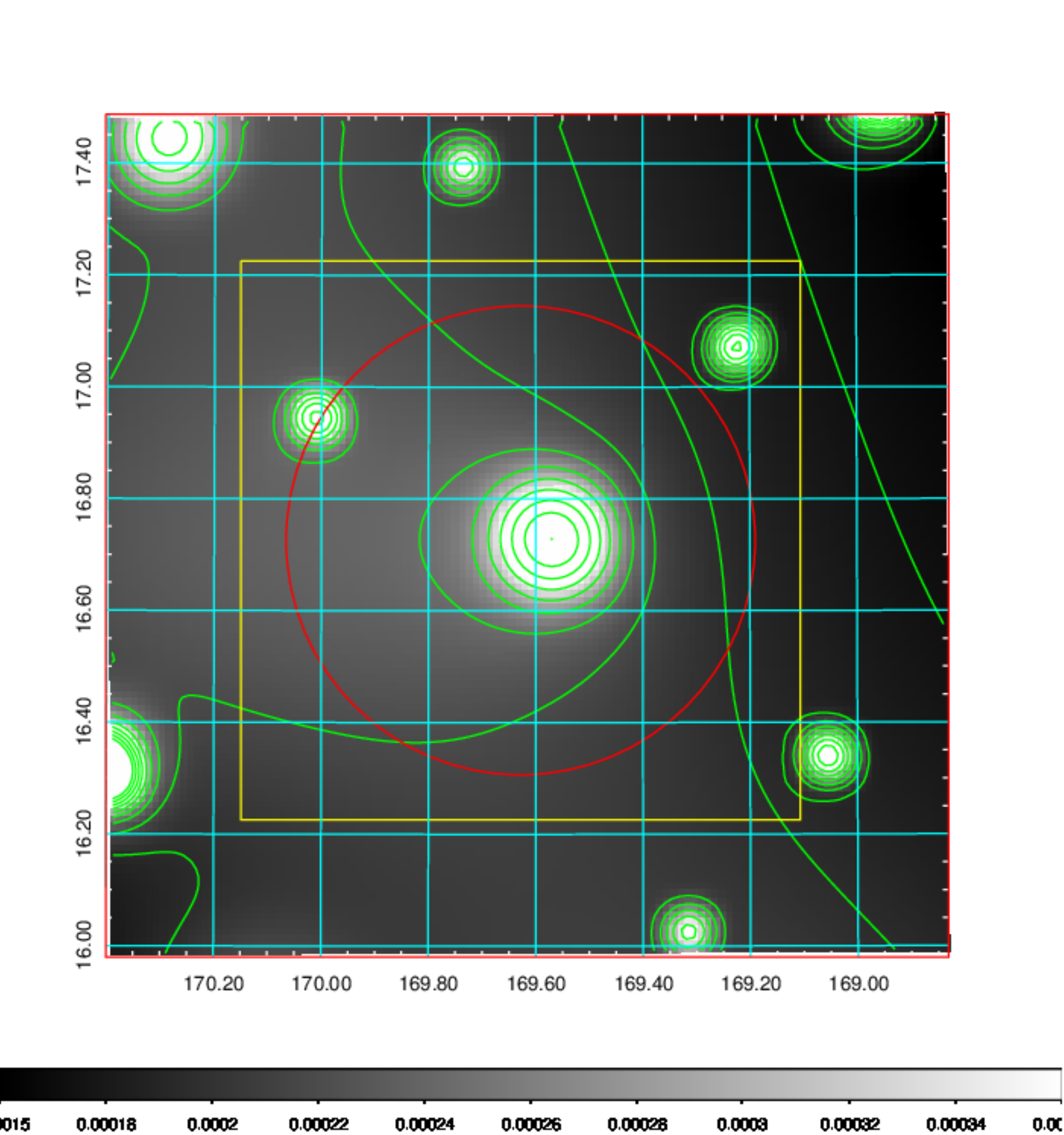} \\ 
\includegraphics[width=0.19\textwidth]{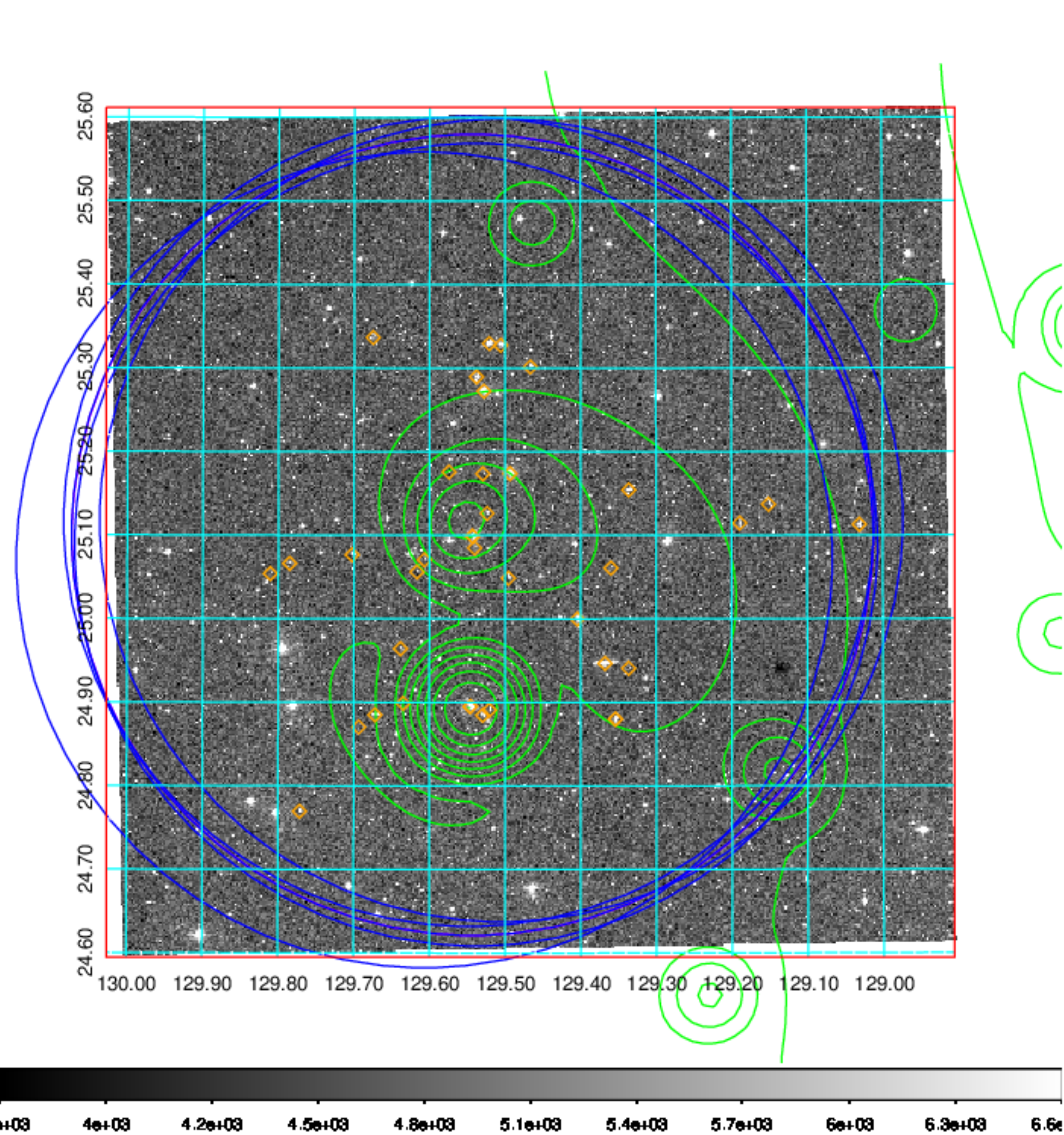} 
\includegraphics[width=0.19\textwidth]{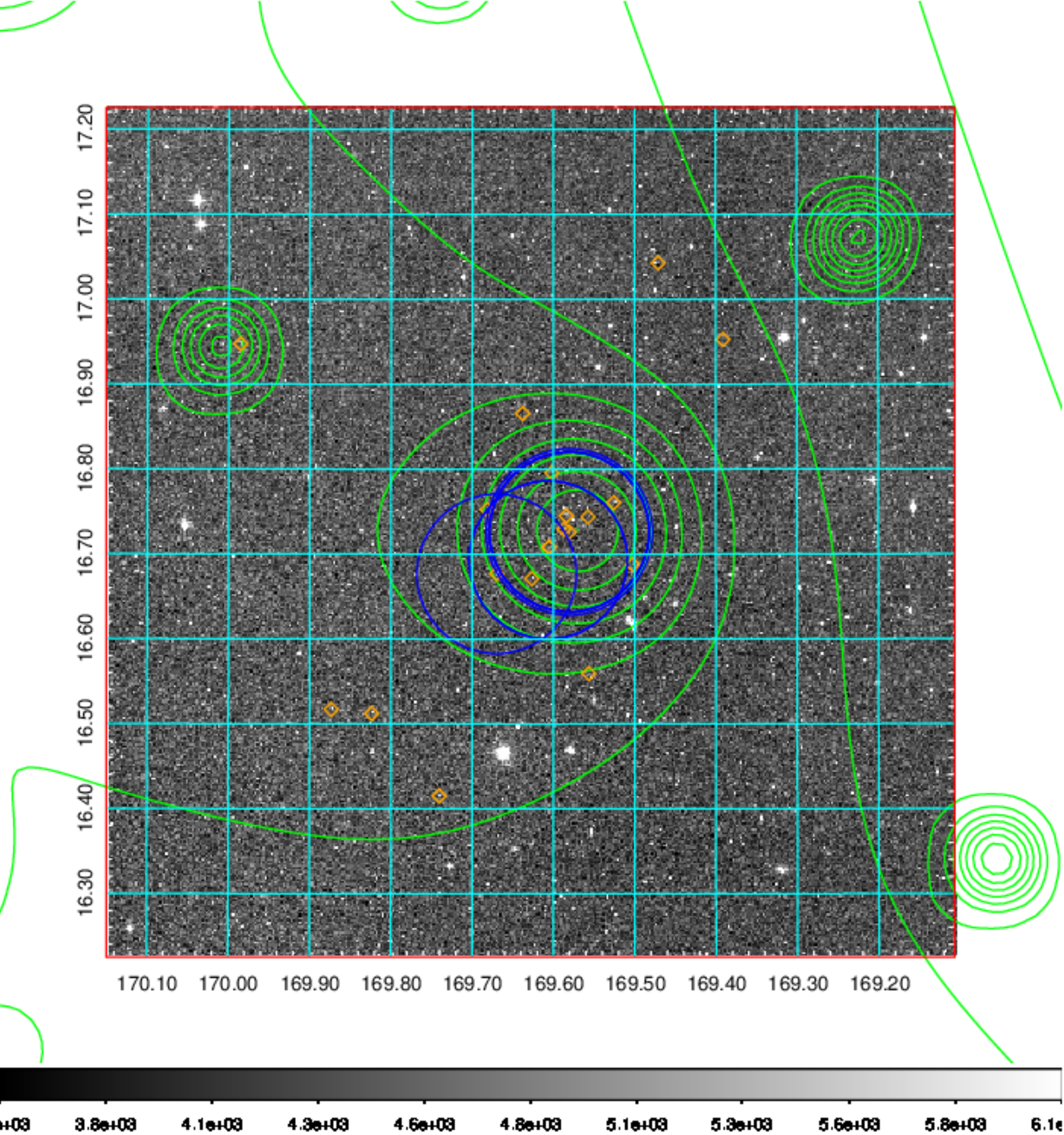} \\ 
\includegraphics[width=0.18\textwidth]{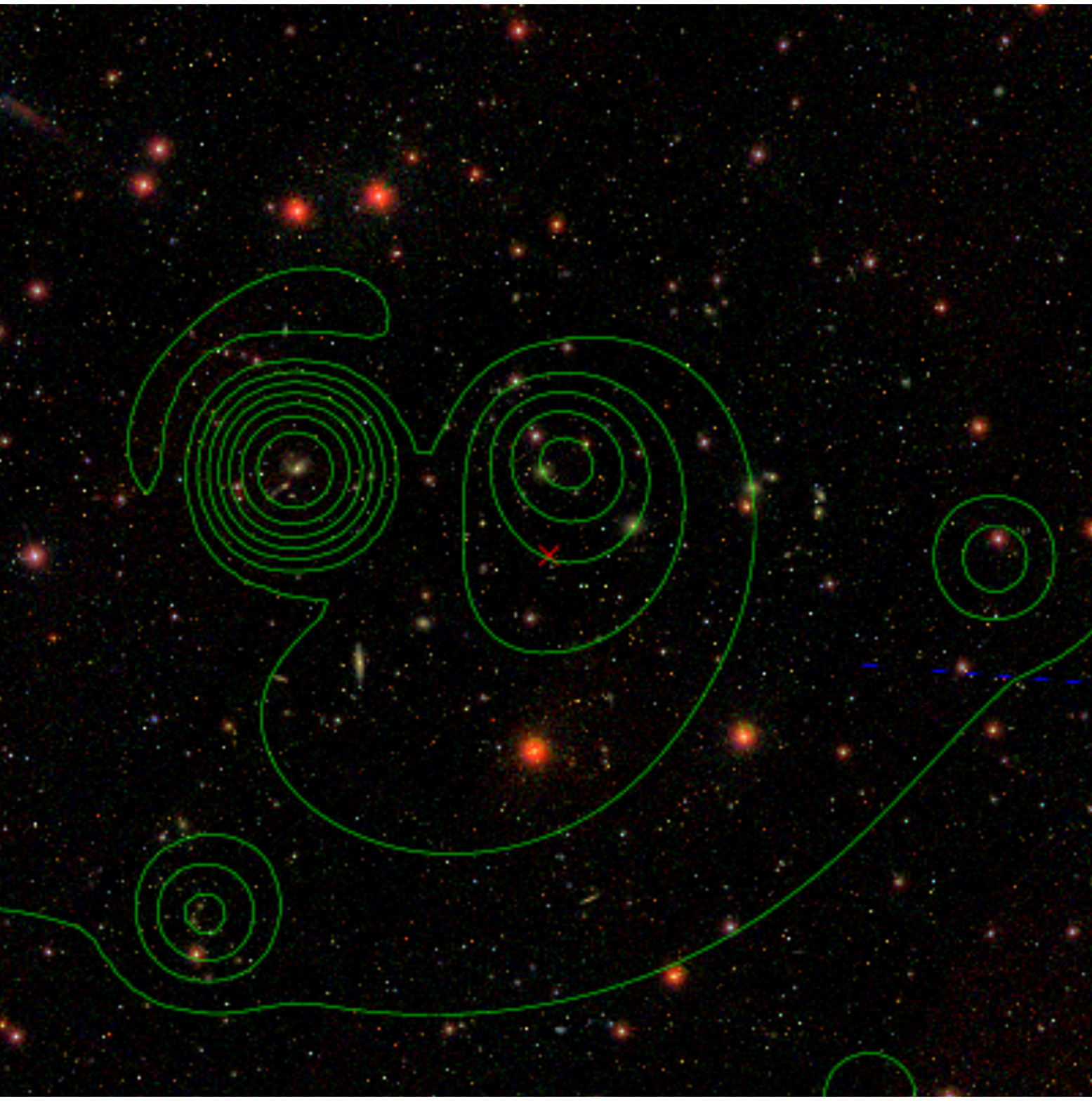} 
\includegraphics[width=0.18\textwidth]{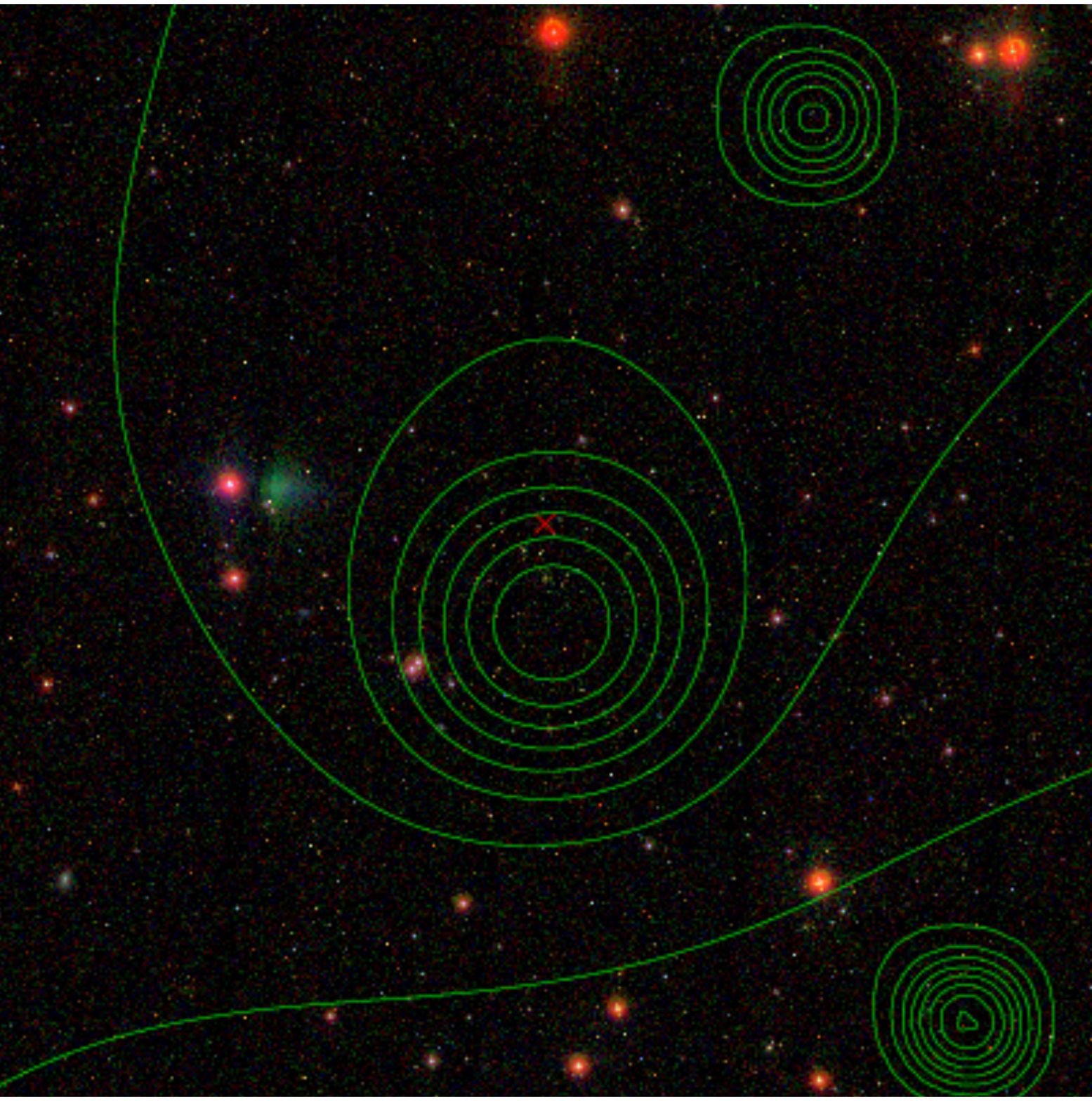} \\ 
\includegraphics[width=0.22\textwidth]{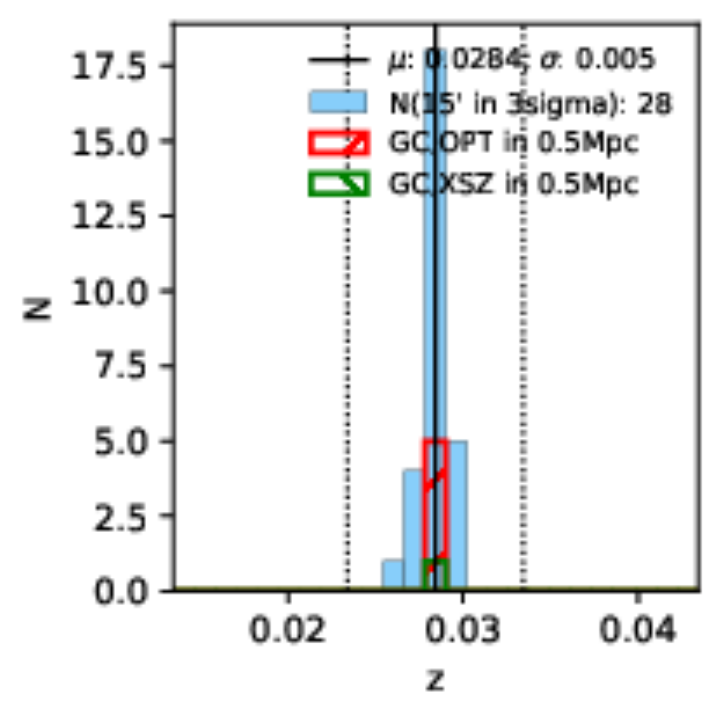} 
\includegraphics[width=0.22\textwidth]{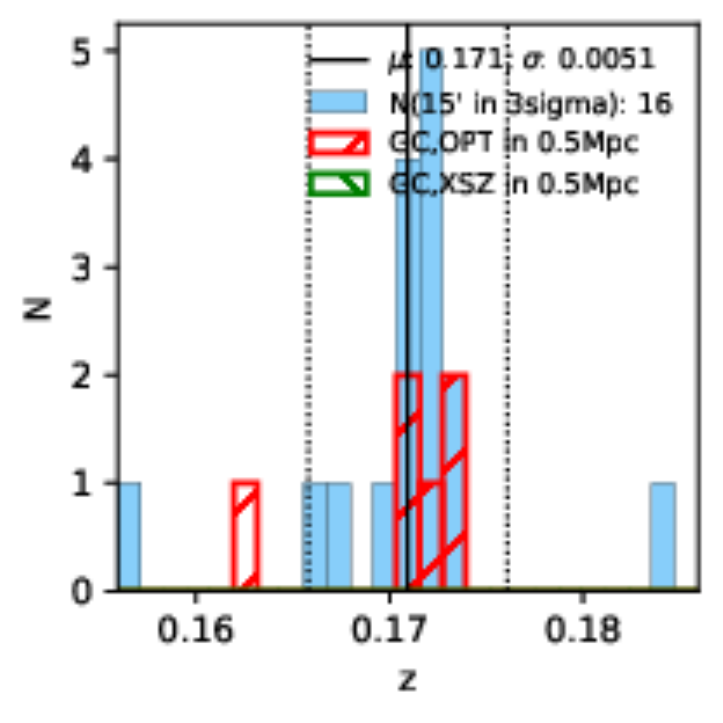} 
\caption{\footnotesize
{Two RXGCC clusters with extent
$>15'$. The left column lists images of RXGCC $296$ and the right column for RXGCC $400$. 
In every column, the panels show, in the sequence of top to bottom, 
the RASS image in [$0.5-2.0$] keV, reconstructed 
image with wavelet filtering, DES image,
SDSS image, and the redshift 
histogram. In the first two rows, the center and radius of red 
circles indicate the position and extent of the candidate, 
while green contours are obtained from the reconstructed image indicating
the smoothed X-ray emission. The images in the third and fourth 
rows are with the size of $1^\circ\times 1^\circ$, which is labeled
with the yellow box in the first two rows. 
The DES images are overlaid with yellow 
diamonds and blue circles for galaxies 
and galaxy clusters with $\Delta z<0.01$, respectively. 
The radius of blue circles is $1~$Mpc. 
The red cross in the fourth panel labels the position 
of our candidate. 
The redshift distribution of galaxies with the offset $<15'$ and $<0.5$~Mpc is
shown in the last row, which is overlaid with the redshift of 
previous optical and ICM-detected clusters respectively in red and 
green histograms. The redshift of the candidate is 
shown in solid black line, while the upper and lower limits of 
1$\sigma$ are shown with black dotted lines. 
}}
\label{fig:example}
\end{figure}

In Fig.~\ref{fig:example}, the images of two such clusters 
are shown as examples. Both of them are with $extent>15'$.
The left column is RXGCC~$296$, and the right column is RXGCC~$400$. 
RXGCC~$296$ is a $Bronze$ cluster, and RXGCC~$400$ is a $Silver$ cluster. 
In each column, the RASS image, reconstructed X-ray image, DES image,
SDSS images, and the redshift histogram of galaxies are shown in the sequence.
As mentioned in Sec.~\ref{subsec:detection}, 
$extent$ is an indication of the size of detected 
clusters, which is shown with the red circle in the 
first two panels of each column.

\section{Redshift and classification change in the catalog}
\label{sec:z_cla_change}

In this section, we add comments about the redshift and classification changes 
for some detections, listed in the Tab.~\ref{tab:app_corz}. The automatic result of the redshift and classification are listed in last two columns, as the method described in Sec.~\ref{subsec:z}, Sec.~\ref{subsec:cluster-literature}, and Sec.~\ref{subsec:class_criteria}. As discussed in Sec.~\ref{subsec:visualinspection}, the visual check resulted in modifications for some candidates, shown in the column $2$ and $3$, with the information of multi-wavelength images, spatial distribution and redshift of galaxies.


\begin{table*}[t]
\centering
\footnotesize
\caption{Redshift or classification changes.}
\begin{threeparttable}
\begin{tabular}{c c c c c | ccccc }
\hline
\hline
RXGCC & $z$ & C &$z$ (auto)&C (auto) &  RXGCC & $z$ & C &$z$ (auto)&C (auto) \\
\hline
 27 & 0.0557 & S &0.0800 & B & 637 & 0.1012 & G &0.0315 & S   \\ 
 48 & 0.2625 & G &0.1537 & S & 644 & 0.0625 & G &0.1349 & B   \\ 
 51 & 0.2300 & B &0.0000 & F & 649 & 0.0312 & S &0.0688 & B   \\ 
 68 & 0.1630 & S &0.0000 & F & 654 & 0.2701 & S &0.0811 & G   \\ 
 86 & 0.0694 & B &0.1235 & B & 659 & 0.2647 & B &0.2957 & G   \\ 
 87 & 0.0170 & S &0.0728 & B & 662 & 0.1668 & B &0.0000 & F   \\ 
 97 & 0.1175 & B &0.0653 & G & 668 & 0.2060 & B &0.0229 & G   \\ 
 99 & 0.3416 & B &0.0000 & F & 695 & 0.0803 & G &0.1109 & G   \\ 
100 & 0.1700 & B &0.0000 & F & 703 & 0.1306 & B &0.0627 & G   \\ 
103 & 0.2198 & B &0.1846 & G & 704 & 0.1603 & B &0.0000 & F   \\ 
111 & 0.2818 & B &0.0000 & F & 706 & 0.1377 & S &0.3383 & G   \\ 
119 & 0.0989 & G &0.0519 & G & 716 & 0.2410 & S &0.1090 & G   \\ 
121 & 0.1290 & G &0.0707 & G & 717 & 0.1035 & S &0.0842 & G   \\ 
131 & 0.2700 & B &0.0000 & F & 718 & 0.2265 & G &0.1089 & G   \\ 
162 & 0.1441 & S &0.0639 & G & 722 & 0.3059 & B &0.0000 & F   \\ 
178 & 0.1039 & B &0.0742 & G & 732 & 0.1406 & B &0.0000 & F   \\ 
182 & 0.2252 & G &0.0000 & F & 734 & 0.1000 & G &0.0466 & G   \\ 
197 & 0.2000 & B &0.0773 & G & 739 & 0.1790 & B &0.0000 & F   \\ 
205 & 0.1516 & G &0.1388 & G & 740 & 0.1410 & B &0.1152 & G   \\ 
207 & 0.2839 & B &0.0330 & S & 742 & 0.1565 & B &0.0000 & F   \\ 
208 & 0.0310 & G &0.0720 & G & 743 & 0.3100 & B &0.0647 & G   \\ 
210 & 0.2977 & G &0.0000 & F & 744 & 0.3280 & B &0.0000 & F   \\ 
220 & 0.1277 & B &0.0515 & S & 750 & 0.2263 & B &0.0586 & G   \\ 
229 & 0.4600 & B &0.0000 & F & 767 & 0.0924 & G &0.0487 & G   \\ 
232 & 0.0977 & G &0.0628 & G & 773 & 0.2764 & B &0.0000 & F   \\ 
237 & 0.4600 & B &0.0000 & F & 774 & 0.1790 & B &0.0000 & F   \\ 
266 & 0.1750 & B &0.0000 & F & 777 & 0.2006 & S &0.0000 & F   \\ 
276 & 0.0222 & B &0.0615 & B & 779 & 0.1080 & B &0.0913 & S   \\ 
306 & 0.4118 & B &0.0897 & G & 782 & 0.1680 & B &0.0162 & G   \\ 
307 & 0.0941 & G &0.0264 & S & 783 & 0.2057 & B &0.0900 & G   \\ 
323 & 0.1381 & S &0.1216 & S & 784 & 0.1623 & B &0.0000 & F   \\ 
326 & 0.1535 & B &0.1272 & G & 797 & 0.0928 & B &0.0715 & S   \\ 
338 & 0.2060 & B &0.1367 & G & 799 & 0.2103 & B &0.0000 & F   \\ 
341 & 0.1193 & G &0.3051 & G & 805 & 0.1886 & B &0.0000 & F   \\ 
346 & 0.1087 & G &0.0206 & S & 807 & 0.2371 & B &0.0000 & F   \\ 
354 & 0.0410 & S &0.1190 & G & 812 & 0.1044 & G &0.0825 & G   \\ 
384 & 0.2480 & G &0.0765 & G & 822 & 0.2746 & B &0.0000 & F   \\ 
401 & 0.1066 & S &0.0705 & B & 833 & 0.1419 & S &0.0593 & G   \\ 
421 & 0.3081 & G &0.0508 & S & 836 & 0.0494 & B &0.0703 & G   \\ 
496 & 0.2953 & S &0.0554 & G & 854 & 0.1402 & G &0.0000 & F   \\ 
570 & 0.2440 & G &0.0986 & G & 919 & 0.0286 & B &0.1867 & S   \\ 
579 & 0.1158 & B &0.1003 & G & 926 & 0.1700 & G &0.0466 & G   \\ 
590 & 0.5551 & B &0.3386 & G & 937 & 0.2478 & G &0.0855 & G   \\ 
596 & 0.2226 & B &0.0000 & F &     &        &   &       &     \\ 
        \hline
        \hline
    \end{tabular}
    \begin{tablenotes}
    \item[Note.]{The $2$nd and $3$rd columns are the estimated redshift and classification listed in the RXGCC catalog. The $4$th and $5$th columns are the redshift and classes obtained with methods in Sec.~\ref{subsec:z}, Sec.~\ref{subsec:cluster-literature}, and Sec.~\ref{subsec:class_criteria}, before the visual inspection discussed in Sec.~\ref{subsec:visualinspection}. }
    \end{tablenotes}
    \small
    \end{threeparttable}
    \label{tab:app_corz}
\end{table*}

\end{document}